\documentclass[12pt,a4paper]{article}
\usepackage{a4}
\usepackage{amsfonts}
\usepackage{amssymb}
\usepackage{epsfig}
\usepackage{amssymb}
\def\beq{\begin{equation}}
\def\be{\begin{equation}}
\def\beqn{\begin{eqnarray}}
\def\ee{\end{equation}}
\def\eeq{\end{equation}}
\def\eeqn{\end{eqnarray}}

\begin{document}
\thispagestyle{empty}

\vspace{1cm}

\begin{center}
{\Large\bf
The Stueckelberg  Z Prime  at the LHC: \\
Discovery Potential, Signature Spaces \\ and Model Discrimination}
\end{center}

\vspace{1.0cm}
\begin{center}

{\bf  Daniel Feldman, Zuowei Liu,  and Pran Nath}

 \vspace{.5cm}

Department of Physics \\
Northeastern University \\
Boston, Massachusetts
\end{center}
\begin{center}
{\bf Abstract} \\
\end{center}
An analysis is given of the capability of the LHC to detect narrow
resonances using  high luminosities  and techniques for
discriminating among models  are discussed.  The analysis is carried
out with focus on the $U(1)_X$ Abelian (Higgless) Stueckelberg
extension of the Standard Model (StSM) gauge group which naturally
leads to a very narrow $Z'$ resonance. Comparison is made to another
class  of models, i.e., models  based on the warped  geometry which
also lead to a narrow resonance via a massive graviton ($G$).
Methods of distinguishing the StSM $Z'$ from the massive  graviton
at the LHC are analyzed using the dilepton final state in the
Drell-Yan process $pp\to Z'\to l^+l^-$ and $pp\to G \to l^+l^-$. It
is shown that the signature spaces in the $\sigma \cdot Br(l^+l^-)
$-resonance mass plane for the $Z$ prime and  for the  massive
graviton are distinct. The angular distributions  in the dilepton
C-M system are also analyzed and it is shown that these
distributions lie high above the background and are distinguishable
from each other. A remarkable result that emerges from the analysis
is the observation that the StSM model with $Z'$ widths even in the
MeV and sub-MeV range for $Z'$ masses extending in the  TeV region
can  produce detectable  cross section signals in the dilepton
channel in the Drell-Yan process with luminosities accessible at the
LHC.   While the result  is derived within the specific StSM class
of models, the capability of the LHC to probe models with narrow
resonances in this range  may hold more generally.\clearpage
\setcounter{footnote}{0}


\section{Introduction}
The Stueckelberg mechanism allows for mass generation of an Abelian
$U(1)$ gauge boson without the benefit of a Higgs mechanism.
Specifically the models of Ref. \cite{Kors:2004dx,kn2,kn3} are based
on the $U(1)$ Stueckelberg extensions of the  Standard Model (SM),
i.e., on the gauge group, $SU(3)_C \times SU(2)_L \times U(1)_Y
\times U(1)_X$. This extension of the SM involves a non-trivial
mixing of the  $U(1)_Y$ hypercharge gauge field $B^{\mu}$ and the
$U(1)_X$  Stueckelberg gauge  field $C^{\mu}$. The Stueckelberg
gauge field $C^{\mu}$ has no couplings with the visible sector
fields, while it  may couple with a hidden sector, and thus the
physical $Z'$ gauge boson connects with the visible sector only via
mixing with the gauge bosons of the physical sector. These  mixings,
however, must be small because of the LEP electroweak constraints
and consequently the couplings of the $Z'$ boson to the visible
matter fields are extra weak, leading to a very narrow $Z'$
resonance. The width of such a boson could be  as  low as a  few MeV
or even lower and lie in the sub-MeV range. An exploration of  the
Stueckelberg $Z'$ boson in the CDF and D\O\ data was  recently
carried  out in Ref. \cite{Feldman:2006ce} and promising prospects
for  its observation at the Tevatron were noted. The models of Ref.
\cite{Kors:2004dx,kn2,kn3} are to be  viewed  as phenomenological,
but may  be low energy effective theories of a more unified
structure. Indeed the Stueckelberg mechanism is quite generic in
string and D brane models
\cite{Dbranes,ghi,Coriano':2005js,Kors:2004iz} but it remains to be
seen if models of the type Ref. \cite{Kors:2004dx,kn2,kn3} can be
embedded in such structures.

The other class of models are those  based on the warped geometry
\cite{Randall:1999ee,Gogberashvili:1998vx} where a narrow massive
graviton excitation with a width lying in tens to hundreds of MeV
can arise in certain regions of its  parameter space.  Thus the
Stueckelberg extensions and the warped geometry models share the
property of allowing for  narrow resonances. It is then pertinent
to investigate the  discovery potential, signature spaces  and model
discrimination for this class of  models  at the LHC.   This is the
main focus of the analysis in this paper. In the first part  of the
paper (Sections 2-7) we will  discuss the discovery potential and
signatures of the  Stueckelberg $Z'$ model. In the second part
(Section 8) we will  carry out a similar analysis  for the case of
warped geometry and present a criteria for model discrimination
between these two classes of models.

\section{A Brief Overview of Stueckelberg Extension of  the SM}
Before proceeding further  we  first  review  the minimal
Stueckelberg extension based  on the gauge group $SU(3)_C \times
SU(2)_L \times U(1)_Y \times U(1)_X$ \cite{Kors:2004dx}. The
effective Lagrangian of  the Stueckelberg extension of the Standard
Model (StSM) can be written as
\begin{equation}
\mathcal{L}_{\rm StSM}=  \mathcal{L}_{\rm St} + \mathcal{L}_{\rm
SM},
\end{equation}
where $\mathcal{L}_{\rm SM} $ is  the Standard Model Lagrangian
\be
\begin{array}{ccc}
&\mathcal{L}_{\rm SM} \supset -\frac{1}{2}{ \rm Tr} \left(F_{\mu \nu
}F^{\mu  \nu }\right)-\frac{1}{4}B_{\mu  \nu }B^{\mu  \nu } +g_2
A_{\mu  }^a\mathcal{J}_{2}^{a \mu }+ g_Y B_{\mu
}\mathcal{J}_{Y}^{\mu  }-\left(  D^{\mu }\Phi \right)^{\dagger }
\left( D_{\mu } \Phi
\right) -V\left( \Phi \right)\\
\end{array}
\ee
 and $\mathcal{L}_{ \rm St}$ is given by
\begin{equation}
\mathcal{L}_{\rm{St}}=-\frac{1}{4}C_{\mu  \nu  }C^{\mu  \nu
}+g_XC_{\mu }\mathcal{J}_X^{\mu }-\frac{1}{2}\left(\partial _{\mu
}\sigma +M_1C_{\mu }+M_2B_{\mu  }\right)^2. \label{StLag}
\end{equation}
Here $C_{\mu}$ is the gauge  field associated with the extra
$U(1)_X$ gauge group and $\mathcal{J}_X^{\mu }$ gives coupling to
the hidden sector but  $C_{\mu}$   has no coupling to the visible
sector; $B_{\mu }$ is the gauge field associated with $U(1)_Y$,
$\sigma$ is the axion,  and $M_1$ and $M_2$ are mass parameters that
appear in the Stueckelberg extension.

\subsection{Mass Matrix of the StSM}\label{MM}
After electroweak symmetry breaking the mass  terms for the neutral
vector bosons take the form
\begin{equation}\mathcal{L}_{\rm StSM}\supset-\frac{1}{2}\mathcal{V}_{\mu} ^ T{M}_{{\rm St}}^2\mathcal{V}^{\mu  },\end{equation}
where
\begin{equation}
\mathcal{ V}^{\mu }=\left(
\begin{array}{c}
 C^{\mu }  \\
 B^{\mu } \\
 A^{3\mu }
\end{array}
\right) \hspace{.5cm}\hspace{.5cm}{M}_{{\rm St}}^2= \left(
\begin{array}{ccc}
    M_1^2             & M_1M_2                                & 0 \\
    M_1M_2         &M_2^2 + \frac{1}{4}v^2g_Y^2  & -\frac{1}{4}v^2g_2g_Y \\
      0                   & -\frac{1}{4}v^2g_2g_Y            & \frac{1}{4}v^2g_2^2
\end{array}
\right),\label{massmatrix}
\end{equation}
and where, $v$ is vacuum expectation value of the Higgs field.
The mass squared matrix, being real and symmetric, can be
diagonalized by an orthogonal transformation
 ${R}^T {M}_{{\rm St}}^2{R}   ={M}_{{\rm St}-\rm {diag}}^2$, with eigenvectors
$ {\mathcal E}^T_{\mu} = (Z'_{\mu}, Z_{\mu}, A^{\gamma}_{\mu})$. The
corresponding eigenvalues, denoted as $\{\lambda_i\}$, are given by
$\{ M_{Z' }^2, M_{Z }^2,M_{\gamma }^2\}$ = $ \{ M_{+ }^2, M_{-
}^2,0\}$ where \beqn
 M_{\pm }^2 &=&
\frac{1}{2} \Bigg [ M^2_0+ M_1^2\left(1+\epsilon ^2\right)
\\
&& \hspace{.5cm} \pm \left[\left({M^2_0}+ M_1^2\left(1+\epsilon
^2\right)\right)^2-4M_1^2\left( M^2_0+ M_W^{2}\epsilon
^2\right)\right]^{1/2}~~ \Bigg] \label{masses} \nonumber, \eeqn and
where \be M^2_0 = \frac{v^2}{4}(g^2_2+g^2_Y), \hspace{.5cm}M^2_W =
\frac{g^2_2 v^2}{4}, \hspace{.5cm} t_W =
\frac{g_Y}{g_2},\hspace{.5cm} \epsilon = \frac{M_2}{M_1} . \ee The
zero eigen-mode is manifest  and is to be associated with the
massless photon state. In the  above model, the  photon field is  a
linear combination of the set of three fields $(C^{\mu}, B^{\mu},
A^{3\mu})$, which is the first indication that the StSM is distinct
from other class of extensions of the SM which predict additonal
spin one gauge bosons
\cite{E6,Hewett:1988xc,Cvetic:1995rj,Babu:1997st,Cho:1998nr,leike,Hill:2002ap,Carena:2004xs,Barger:2006dh}.
In the limit $M_2 \ll M_1 $, i.e. $ \epsilon \to 0,$ the
Stueckelberg sector decouples from the Standard Model and the tree
level expressions for the Standard Model  $Z$  boson mass is
recovered, while the ${Z'}$ mass limits  to $ M_1$ which is the
overall scale of new physics in the StSM.
As discussed above, the physical fields $ {\mathcal E}^T_{\mu} =
(Z'_{\mu}, Z_{\mu}, A^{\gamma}_{\mu})$ are related to the fields
${\mathcal V}^T_{\mu} = (C_{\mu}, B_{\mu}, A^3_{ \mu})$ through the
orthogonal transformation ${\mathcal V}_{\mu} = R {\mathcal
E}_{\mu}$. The matrix  $R$ is easily formed from the eigenvectors
$\xi_{\lambda_i}$ so that one may write $R =(
\xi_{\lambda_1},\xi_{\lambda_2},\xi_{\lambda_3})$, where \be
{\xi}_{\lambda_i} =\Bigg[\left(\frac{{M^2_1} \epsilon
}{-M^2_W{t_W}}\frac{M^2_W-{\lambda_i}}{{M^2_1}-{\lambda_i}}
\right)^2+\left(\frac{M^2_W-{\lambda_i}}{M^2_W{t_W}}\right)^2+1\Bigg]^{-1/2}
\left(
\begin{array}{c}
 \frac{{M^2_1} \epsilon }{-M^2_W{t_W}}\frac{M^2_W-{\lambda_i}}{{M^2_1}-{\lambda_i}} \\
 \frac{M^2_W-{\lambda_i}}{M^2_W{t_W}} \\
 1
\end{array} \right)
\label{eig} \ee and where $\{\lambda_i\}$ are the eigenvalues of the
mass matrix of Eq. (\ref{massmatrix}) as given above.
\subsection{Neutral Current Interactions of the StSM}\label{StVS}
\begin{figure}[h]
\vspace{3 cm}
\begin{center}
\includegraphics[width=14cm,height=12cm]{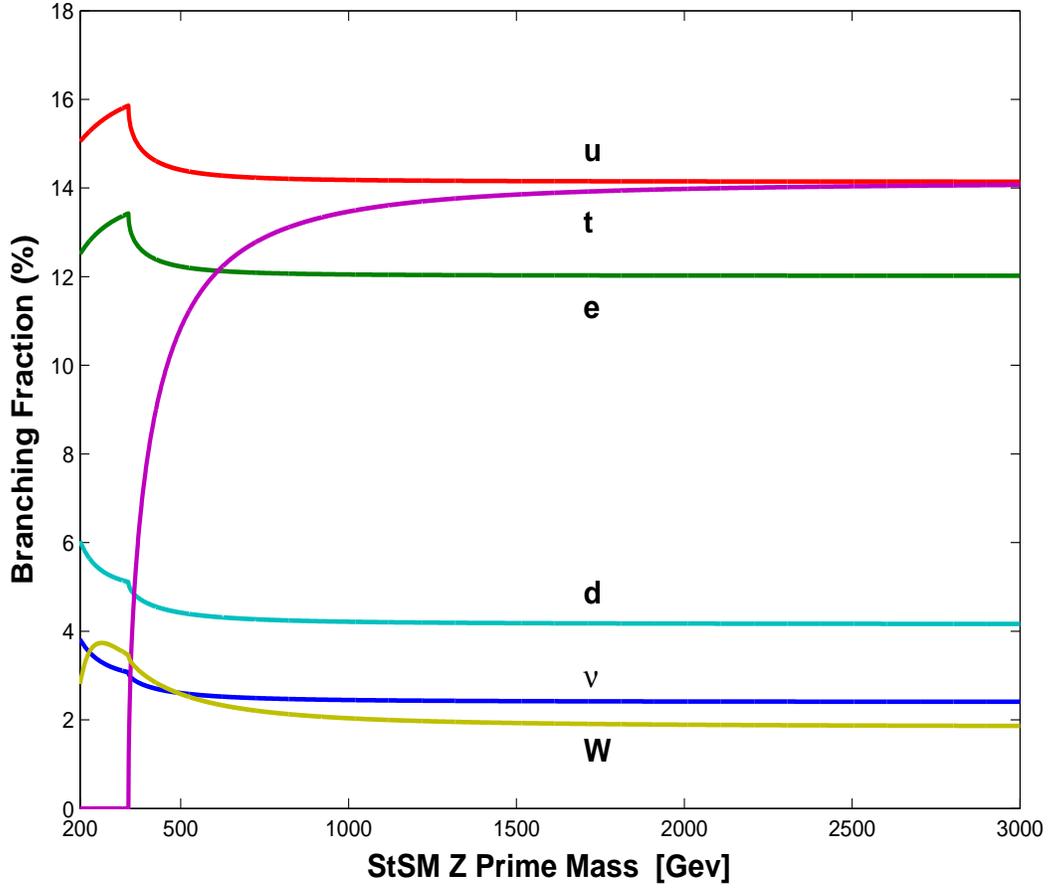}
\end{center}
\caption{ The StSM $Z'$ branching ratios into $f\bar f$ and $W^+W^-$
final states as a function of the $Z'$ mass with $f=u,t,e,d,\nu$
with $\epsilon=0.06$. Besides the exceptionally narrow total decay
width, the large  branching ratio of the StSM $Z'$ into charged
leptons further distinguishes this model from other $Z'$ models.}
\label{fig:StSMBr}
\end{figure}

The interaction Lagrangian in the neutral sector of the StSM,
involving the couplings of visible matter to  the gauge fields, is
given by \be {\mathcal L }_{N} =  g_M \sum\limits_{f } {\bar f}
\gamma^{\mu} [ (v_f- \gamma_5 a_f) {Z}_{\mu} + (v'_f- \gamma_5 a'_f)
Z'_{\mu}]f +
 e A^{\gamma}_{\mu }\left(\mathcal{J}_{Y}^{\mu }+ \mathcal{J}_{2L}^{3
 \mu}\right).
\ee
Here $g_M = (\sqrt{ 2} G_F M^{2}_Z)^{1/2} = 2/\sqrt{g^2_2+g_Y^2}$,
and the electrical charge $e$ is given by \be \frac{1}{e^2} =
\frac{1}{g^2_2}+ \frac{1}{g^2_Y} (1+\epsilon^2) \ee where $e$ limits
to the SM relation as $\epsilon \to 0$. The couplings to the $Z$ and
$Z'$ gauge bosons are then determined to be \be
\begin{array}{l}
v_f  = (c_W R_{3 2} - s_W R_{2 2})T^{3 }_{f } + 2 Q_f s_W R_{2 2} \\
a_f =  (c_W R_{3 2} - s_W R_{2 2})T^{3 }_{f },   \\
 \end{array}\label{Zcouplings}
\ee \be
\begin{array}{l}
v'_f =(c_W R_{3 1} - s_W R_{2 1})T^{3 }_{f }  + 2 Q_f s_W  R_{2 1}  \\
a'_f = (c_W R_{3 1} - s_W R_{2 1})T^{3 }_{f },  \\
 \end{array}\label{StSM-coup}
\ee where $c_W= g_2/\sqrt{g^2_2+g_Y^2}$, and $s_W =g_Y/
\sqrt{g^2_2+g_Y^2}$. In the  limit ${\epsilon \to 0}$ one has  $R_{3
1},R_{2 1}, \to 0$,  $R_{2 2}\to -s_W$ and $ R_{3 2} \to c_W$ (see
Eqs. (\ref{eig})),
 so that
 $v_f\to v_f(SM)  =T^{3 }_{f } - 2 Q_f s^2_W$ and
 $a_f\to a_f(SM)=T^{3 }_{f }$.
The coupling structure of the Stueckelberg $Z'$ gauge boson with
visible matter fields is suppressed by small mass mixing parameters
 thus leading to a very narrow $Z'$ resonance. As will be discussed in Sections (6-8), such a resonance may be detectable
 via the  Drell-Yan process at the LHC by an analysis of a dilepton pair arising
from the decay of the $Z'$.
The partial fermion decay widths of the StSM $Z'$ are given by
\begin{eqnarray}
\Gamma(Z' \to \nu \bar \nu)&=& \frac{G_F M_Z^2}{6\sqrt{2}\pi}M_{Z'}\left[v'^{2}_{\nu}+a'^{2}_{\nu}\right] \\
\Gamma(Z' \to e \bar e)     &=& \frac{G_F M_Z^2}{6\sqrt{2}\pi}M_{Z'}\left[v'^{2}_{e}+a'^{2}_{e}\right] \\
\Gamma(Z' \to u \bar u)    &=& N_c \frac{G_F M_Z^2}{6\sqrt{2}\pi}M_{Z'}\left[v'^{2}_{u}+a'^{2}_{u}\right]\left(1+\frac{\alpha_{s}}{\pi}\right)\\
\Gamma(Z' \to d \bar d)    &=& N_c \frac{G_F M_Z^2}{6\sqrt{2}\pi}M_{Z'}\left[v'^{2}_{d}+a'^{2}_{d}\right]\left(1+\frac{\alpha_{s}}{\pi}\right)\\
\Gamma(Z' \to t \bar t)    &=& \theta(M_{Z'}-2m_{t})
N_c\frac{G_FM_Z^2}{6\sqrt{2}\pi}M_{Z'}\sqrt{1-\left(\frac{2m_{t}}{M_{Z'}}\right)^2}\\\nonumber
&&\times\left[v'^{2}_{t}\left(1+2\frac{m_t^2}{M_{Z'}^2}\right)+a'^{2}_{t}\left(1-4\frac{m_t^2}{M_{Z'}^2}\right)\right]
\left(1+\frac{\alpha_{s}}{\pi}\right),
\end{eqnarray}
where $N_c =3$ and we have included the leading order QCD
corrections, but neglected the relatively small electroweak
corrections and fermion masses except for the top quark mass.
Additionally for $M_{Z'}>2M_W$, the $Z'$ can decay into $W^+W^-$
which is determined by the triple gauge boson vertex,
\begin{equation}
\mathcal{L}_{Z'WW}=i g_2 R_{31}\left[W_{\mu \nu
}^+W^{-\mu}Z'^{\nu}+W_{\mu \nu }^-W^{+\nu }Z'^{\mu}+W^{+\mu
}W^{-\nu}Z'_{\mu \nu}\right].
\end{equation}
The $W^+W^-$ decay width is  then given by \beqn
\Gamma(Z'\rightarrow W^+ W^-) = \theta(M_{Z'}-2M_W) \frac{g_2^2
R_{31}^2}{192\pi}M_{Z'}\frac{M_{Z'}^4}{M_W^4}
\left[1-4\frac{M_W^2}{M_{Z'}^2}\right]^{\frac{3}{2}}\nonumber\\
\times
\left[1+20\frac{M_W^2}{M_{Z'}^2}+12\frac{M_W^4}{M_{Z'}^4}\right],
\eeqn in agreement with previous analyses of $Z'$ decays
\cite{Barger:1987xw,Dutta:1993ee}. The $W^+W^-$ decay mode is
suppressed by the small factor $R_{31}$, the element of the rotation
matrix which indicates the mixing between $Z'$ and $A^3$ gauge
bosons.
 The  $\Gamma( Z' \rightarrow W^+W^-)$
 width is typically small relative to $\Gamma( Z' \rightarrow \sum_i f_i\bar f_i)$.
It will be shown in the following sections that $\epsilon$ is
severely limited by the electroweak constraints which leads to a
Stueckelberg $Z'$ resonance with a very narrow decay width.    Thus
the $Z'$ decay width lies  in the $\leq 100$  MeV range with
$M_{Z'}$ lying in the several hundred GeV to $1~\rm TeV$ range. In
Fig. (\ref{fig:StSMBr}) it is shown that  the  $Z'$   decays into
quarks and leptons will
 dominate the total $Z'$ decay width, as the $W^+W^-$  decay mode is roughly the same size as one species
of $\nu \bar \nu$ mode. One may note  that the branching ratio of
$Z'$  into the charged  leptons is relatively large  compared  to
what  one has  in conventional models. This is due to the StSM $Z'$
couplings being  dominated  by the hypercharge  of the particle in
the final state. Thus, the isospin singlet  $l_R$ which has a
hypercharge $Y=-2$ contributes a  significant amount which makes the
charged lepton contribution comparable to the up quark contribution
overcoming the color factor. The above  also indicates  that this
$Z'$ model can be efficiently tested in an $e^+e^-$ collider with
polarized beams where one could check on the $l_R$ vs  $l_L$
couplings. Such an experiment will be possible at the  ILC.
 The above, coupled with the Drell-Yan analysis  is   a prime example of the physics interplay between the ILC and LHC  \cite{Weiglein:2004hn}.

\section{The Stueckelberg Extension of LR Symmetric Models}
\subsection{Mass Matrix and Interactions}
Next we discuss  the Stueckelberg extension of the
 Left-Right Symmetric model (abbreviated by StLR) introduced in \cite{Feldman:2006ce}. The gauge sector  of this  group
 is given by $SU(2)_L\times SU(2)_R\times U(1)_{B-L}\times U(1)_X$ with gauge bosons $A^{a \mu}_L, A^{a \mu}_R, B^{\mu}, C^{\mu}$.
As in LR models  we  assume the Higgs  sector  of the model to
include  $SU(2)_L$ and $SU(2)_R$  doublets $\Phi_{L,R}$ and a
$SU(2)_L\times SU(2)_R$ bi-doublet $\xi$. We  take the Lagrangian
for the  extended  model to be
 \be
\mathcal{L}_{\rm StLR} =   \mathcal{L}_{\rm St} + \mathcal{L}_{\rm
LR} ,\ee
where $\mathcal{L}_{\rm St}$ is the same as in StSM and is  given by
Eq. (\ref{StLag}), and where $\mathcal{L}_{\rm LR} $  is the
standard  Left Right Symmetric Lagrangian \cite{Mohapatra:1974gc}
which we  display below to define notation
\begin{eqnarray}
\mathcal{L}_{\rm LR} &=& -\frac{1}{2}{ \rm Tr} \left(F_{L\mu
\nu}F_L^{\mu  \nu }\right) -\frac{1}{2}{ \rm Tr} \left(F_{R\mu
\nu}F_R^{\mu  \nu }\right) -\frac{1}{4}B_{\mu  \nu  }B^{\mu \nu }
\\ \nonumber &&+g A_{L \mu}^a \mathcal{J}_{2L}^{a \mu } +g A_{R \mu
}^a \mathcal{J}_{2R}^{a \mu } + g' B_{\mu }\mathcal{J}_{B
-L}^{\mu}-\left(D^{\mu }\Phi _L\right)^{\dagger }D_{\mu }\Phi _L
      \\ \nonumber
&&-\left(D^{\mu}\Phi _R\right)^{\dagger }D_{\mu }\Phi _R-{\rm
Tr}\left[\left(D^{\mu }\xi \right)^{\dagger }\left(D_{\mu }\xi
\right)\right]-V\left(\Phi_L,\Phi _R,\xi \right).
\end{eqnarray}
We work with the manifest L-R symmetry $g=g_{2L}=g_{2R}$, and we use
the notation $g' = g_{BL}$.
The set of Higgs multiplets under one pattern of symmetry breaking
takes the form $\langle \Phi_L \rangle = v_L/\sqrt{2}$, $\langle
\Phi_R \rangle = v_R/\sqrt{2}$, and
\begin{equation}
 \langle \xi \rangle=\frac{1}{\sqrt{2}} \left(\begin{array}{cc}
 \kappa  & 0 \\
 0 & \kappa '
\end{array}
\right),
\end{equation}
with $\kappa ' \ll \kappa \ll v_R$, $v_L v_R = \gamma \kappa^2$
 and $\gamma$ being the ratio of Higgs-particle self-couplings  \cite{Mohapatra:1974gc}.
The mass squared matrix in the neutral sector is given by
\be {M}_{{ \rm StLR}}^2=\left(
\begin{array}{cccc}
 M_1^2 & M_1M_2 & 0 & 0 \\
 M_1M_2 & \frac{1}{4}\left(v_L^2+v_R^2\right)g'^2+M_2^2 & -\frac{1}{4}g g'v_L^2 & -\frac{1}{4}g g'v_R^2 \\
 0 & -\frac{1}{4}g g'v_L^2 & \frac{1}{4}g^2\left(v_L^2+\kappa ^2+\kappa '^2\right) & -\frac{1}{4}g^2\left(\kappa ^2+\kappa '^2\right) \\
 0 & -\frac{1}{4}g g'v_R^2 & -\frac{1}{4}g^2\left(\kappa ^2+\kappa '^2\right) & \frac{1}{4}g^2\left(v_R^2+\kappa ^2+\kappa '^2\right)
\end{array}
\right) \label{masslr}\ee
which enters in the Lagrangian through
 \be \mathcal{L}_{{
\rm StLR}}\supset-\frac{1}{2} \tilde{\mathcal{V}}_{\mu}^{  T} {M}_{{
\rm StLR}}^2 \tilde{\mathcal{V}}^{\mu  } \hspace{.5cm}{ \rm
with}\hspace{.5cm} \tilde{\mathcal{V}}_{\mu}^{  T} =
 (C_{\mu },B_{\mu },  A_{L \mu}^3,  A_{R \mu}^3).
\ee
The matrix of Eq. (\ref{masslr}) contains a massless mode, i.e. the
photon, and  three massive modes $Z, Z', Z''$.
 We arrange the eigenvalues of ${M}_{{ \rm StLR}}^2$ in the order
\be{M}_{{ \rm StLR-diag}}^2={ \rm
diag}\left(M_{Z'}^2,M_Z^2,0,M_{Z''}^2\right),\ee
with the corresponding eigenvectors \be \tilde{
\mathcal{E}}^T_{\mu}= \left(
 Z'_{\mu },
 Z_{\mu },
 A^{\gamma}_{\mu },
 Z'' _{\mu }
\right),\ee where $ \tilde{ \mathcal{V}}^{\mu} $ and $
\tilde{\mathcal{E}}^{\mu }$ are related by $\tilde{\mathcal{V}}^{\mu
}=\mathcal{O} \tilde{\mathcal{E}}^{\mu }$, where $\mathcal O $ is an
orthogonal matrix, $\mathcal O^T\mathcal O=I$. In our notation
$A^{\gamma}_{\mu}, Z_{\mu}, Z_{\mu}^{''}$ are the usual modes in the
LR model and $Z'_{\mu}$ is the new mode arising due to mixing with
the Stueckelberg sector.  In this model the neutral current
interactions have the form
\begin{equation}
g_M \sum\limits_{f } {\bar f} \gamma^{\mu} [ (v_f- \gamma_5 a_f)
{Z}_{\mu} + (v'_f- \gamma_5 a'_f) Z'_{\mu}]f +e A_{\mu
}^{\gamma}\left(\mathcal{J}_{B-L}^{\mu}+\mathcal{J}_{2L}^{3\mu }+
\mathcal{J}_{2R}^{3 \mu }\right)
\end{equation}
where $e$ is given by \be \frac{1}{e^2} =\frac{1}{g^2}
(1-\epsilon^2) + \frac{1}{g_{Y}^2}  (1+\epsilon^2) \ee
 and where $g_Y$  is related to $g=g_{2L} =g_{2R}$ and $g_{BL}=g'$  by
$ {1}/{g_Y^2} ={1}/{g^2}+{1}/{g^2_{BL}}$. The above relations limit
to the standard LR relation as
 $\epsilon = {M_2}/{M_1}\to 0 $.\\

The vector and axial vector couplings  of $Z$ and $Z'$ to the matter
fields are determined as in Section \ref{StVS} and are,

\be
\begin{array}{l}
 v_f     =   \frac{1}{\sqrt{g_2^2+g_{Y}^2}}
                        [ g ({\mathcal{O}}_{32}+{\mathcal{O}}_{42}) T^3_f + g' {\mathcal{O}}_{22} (B-L)_f] \\
 a_f     =    \frac{1}{\sqrt{g_2^2+g_{Y}^2}}
                       [ g ({\mathcal{O}}_{32}-{\mathcal{O}}_{42}) T^3_f ],   \\
 \end{array}\label{StLR-Zcoup}
\ee \be
\begin{array}{l}
v'_f    =   \frac{1}{\sqrt{g_2^2+g_{Y}^2}}
                       [ g ({\mathcal{O}}_{31}+{\mathcal{O}}_{41}) T^3_f + g' {\mathcal{O}}_{21} (B-L)_f ]  \\
  a'_f   =    \frac{1}{\sqrt{g_2^2+g_{Y}^2}}
                       [ g ({\mathcal{O}}_{31}-{\mathcal{O}}_{41}) T^3_f ]. \\
 \end{array}\label{StLR-coup}
\ee
The StLR $Z'$ and StSM $Z'$ share remarkably similar properties. A
comparison between these two models is exhibited in Table (2). The
analysis shows the interesting phenomenon that although the maximum
allowed value of $\epsilon$ in the StLR is somewhat larger than in
the StSM, the constraints on the axial-vector and vector couplings
of  the $Z'$ with quarks and leptons and  on the couplings with
$W^+W^-$   are very similar to those in StSM. Consequently the
branching ratios of the $Z'$ into these  modes are very similar.
Thus as in the case of the StSM, one also finds that in the StLR,
the dominant contribution to the decay of the $Z'$  is from the
quark and lepton  final states. Restrictions on the parameter space
of the limiting form of the StLR, which is the LR model, show that
the decay into the extra heavy $W^+W^-$ final state is not
kinematically allowed.

\section{Constraints on the $U(1)_X$ Extensions}\label{Ewconst.}
\subsection{Constraint from the Correction to the $Z$ Mass }\label{ConstZmass}
We use the variational technique of Ref. \cite{Arnowitt:1992qp} to
derive the shift on the $Z$ mass  due to the effect of mixing with
$C_{\mu}$. In general, for a real symmetric $n \times n$ matrix, the
eigenvalue equation is an $n^{\rm th}$
 order polynomial in $\lambda$
  \be F(\lambda ) = \sum\limits_{k = 1}^n {C^{(k)} }  {\lambda}^k  = 0.
  \ee
The correction to an eigenvalue $\lambda_i$  due to a set of
perturbation $\delta _k$ may be written as \be\Delta \lambda _i  =
\sum\limits_{k = 1}^m {\delta _k \frac{{\partial \lambda _i
}}{{\partial \delta _k }} }  =  - \sum\limits_{k = 1}^m {\delta _k
\left( {\frac{{\partial _{\delta _k } F}}{{\partial _\lambda  F}}}
\right)_{\lambda  = \lambda^* _{ik} } }\label{vari}, \ee where
$\lambda^{*}_{ik}  = \lim_{\delta_k \to 0}\lambda_i$. For the
$U(1)_X$ extended theory we have after factoring out the  zero
eigenvalue the equation $ F(\lambda) = C^{(2)}  \lambda^2 +C^{(1)}
\lambda+C^{(0)} $ with \be
\begin{array}{l}
C^{(2)}  =  1 \\
C^{(1)}  =  -(M^2_0+M^2_1+M^2_2) \\
C^{(0)}  =   M^2_1 M^2_0 + M^2_0  M^2_2 c^2_W,
\end{array}
\ee where we are  interested in the shift on the $Z$ mass (as given
by  Eq. (\ref{masses}))  due to the perturbation $\delta = M^2_2 $.
The above gives

\be\Delta M_Z \approx -\frac{1}{2} M_0 s^2_W(1-{M_0^2}/{M_1^2})^{-1}
\epsilon^2. \label{shift}\ee
To determine  the  allowed corridors  in $\epsilon$ and $M_1$, we
follow a similar approach as  in the analysis of Refs.
\cite{Nath:1999fs,marciano}   used in constraining  the size of
extra dimensions.
  We begin by recalling that in the
 on-shell scheme the  $W$  boson mass including loop corrections becomes \cite{Sirlin:1983ys}
  \be
 {M}_W^2 \to \frac{\pi \alpha}{ \sqrt 2 G_F s^2_W (1-\Delta r)}  ,
 \label{wmass}
  \ee
 where the Fermi constant $G_F$ and the
fine structure constant $\alpha$  (at  $Q^2=0$)  are  known to a
high degree  of accuracy. The quantity $\Delta r$ is the radiative
correction and is determined so that
 $\Delta r= 0.0363\pm 0.0019$ \cite{:2005em}, where the  uncertainty comes from error in the top mass and
 from  the error in $\alpha (M_Z^2)$.    Since in the on-shell
 scheme $s^2_W= (1-M_W^2/M_Z^2)$ one may use Eq. (\ref{wmass})  and the current experimental
 value of $M_W=80.425 \pm 0.034$ \cite{:2005em}  to make a prediction of
 $M_Z$.  Such a prediction within the SM  is in excellent agreement with the
 current experimental value of $M_Z=91.1876\pm 0.0021$.
 Thus the above analysis requires that the effects of the Stueckelberg extension on the $Z$ mass must be
 such that they lie in the error corridor of the SM  prediction.
 From Eq.  (\ref{wmass}) we find
  \be
\delta M_Z  =M_Z\sqrt{
\left(\frac{1-2\sin^2\theta_W}{\cos^3\theta_W} \frac{\delta
M_W}{M_Z}\right )^2
 +\frac{\tan^4\theta_W (\delta \Delta r)^2}{ 4(1-\Delta r)^2 }  }.
 \label{delta1}
 \ee
Equating the StSM shift of the $Z$ mass, Eq.  (\ref{shift}), in the
region $M_1^2 \gg M_Z^2$,  to the SM error corrider of the $Z$ mass,
Eq. (\ref{delta1}),
 one finds
 an {\it upper bound}  on $\epsilon$ \cite{Feldman:2006ce}
 \beqn
 |\epsilon| \lesssim
  .061 \sqrt{ 1-(M_Z/M_1)^2}
 \label{epsconstraint}.
 \eeqn

\subsection{Constraints from Other Precision Electroweak Data}
 Next we investigate the implications of the previous analysis on the
  precisely determined observables in the electroweak sector.
  We follow closely the analysis of the LEP Working Group  \cite{:2005em} (see also Refs.  \cite{Baur:2001ze,Bardin:1999gt}),
  except that we will use the vector ($v_f$) and the axial vector ($a_f$) couplings  for the fermions in the StSM.
  The couplings  of the
$Z$  to the fermions in the StSM are elevated from the tree level
expressions of Eqs. (\ref{Zcouplings}) to \be
\begin{array}{l}
v_f  =\sqrt{\rho_f}[ (c_W R_{3 2} - s_W R_{2 2})T^{3 }_{f } + 2 \kappa_f Q_f s_W R_{2 2} ] \\
a_f= \sqrt{\rho_f}  (c_W R_{3 2} - s_W R_{2 2})T^{3 }_{f }, \\
 \end{array}
\ee
 where $\rho_{f}$ and $\kappa_{f}$ (in general complex valued quantities) contain radiative corrections from propagator self energies and flavor specific vertex corrections and are as defined in Refs. \cite{Erler:2004nh,:2005em}.
 The decay of the $Z$ boson
   into  lepton anti-lepton and quark anti-quark pairs (excluding the top) in the on-shell renormalization scheme is given by
   \cite{Baur:2001ze,Erler:2004nh}
\beqn
 \Gamma(Z\to f\bar f) &=& N_f^c  \mathcal{R}_f \Gamma _o  \sqrt {1-4\mu _f^2 } \Bigg  [ |v_f|^2 (1 + 2\mu _f^2 ) + |a_f|^2 (1 -4\mu _f^2 )\Bigg], \\
 \mathcal{R}_f  &=& \left(1 + \delta _f^{QED} \right)\left(1 +\frac{N_f^c-1}{2}\delta _f^{QCD} \right),\\
 \delta _{f}^{QED}  &=&\frac{3\alpha}{4\pi}Q_f^2 , \\
 \delta _{f}^{QCD}  &=& \frac{{\alpha_s }}{\pi } + 1.409\left( {\frac{{\alpha _s }}{\pi }} \right)^2
 -12.77\left( {\frac{{\alpha _s }}{\pi }} \right)^3  - Q_f^2
\frac{{\alpha \alpha _s   }}{{4\pi ^2 }}.
 \eeqn
Here  $\alpha $ and  $\alpha _s $ are taken at the $M_Z$ scale,
 while $ N_f^c =(1,3)$ for leptons and quarks. In the above, $ \Gamma_o =G_F M^3_Z/ 6\sqrt 2 \pi$, and  $\mu _f = m_f  / M_Z$.
The total decay width ($\Gamma_Z$) of the $Z$ into quarks and
leptons, in the visible sector,
 is just the sum over all the final states.\\

 We also investigate the effects of mixing with the Stueckelberg sector  on the following $Z$ pole
 observables
\begin{eqnarray}
R_{l}       &=&    \frac{\Gamma(had)}{\Gamma(l^+l^-)},     \\
R_{q}       &=&    \frac{\Gamma(q\bar q)}{\Gamma(had)},   \\
\sigma_{had}&=&    \frac{12\pi \Gamma(e^+e^-)\Gamma(had)}{M_Z^2 \Gamma_Z^2},  \\
A_{f}       &=&    \frac{2 v_f a_f}{v^2_f+a^2_f}, \\
A^{(0,f)}_{FB}  &=&         \frac{3}{4} A_e A_f.
\end{eqnarray}
Using the above we have carried out a fit in the electroweak sector
on  the quantities  sensitive to mixing with the Stueckelberg
sector.  A summary of the analysis is presented in   Table
(\ref{tab:table1}) for
 $M_1=350$ GeV and $\epsilon$ lying in the range (0.035-0.059). The analysis of Pulls  in Table (\ref{tab:table1})
 indicates that the fits are  excellent.  Indeed for the case $\epsilon =.035$, the StSM gives essentially the
 same  $\chi^2$ fit to data as the SM.  For the case $\epsilon =0.059$ the Pulls  are again of the same quality
 as  for the SM when $A_{FB}^{(0,b)}$ is  excluded but somewhat larger when $A_{FB}^{(0,b)}$ is included.
 However,  $A_{FB}^{(0,b)}$  is known to be  problematic even in the SM.
  Thus, for example, $A_{FB}^{(0,b)}$  lies  in the range [-2.5,-2.8]  in the analysis of Ref. \cite{:2005em}  and it is implied that
 the significant shift could be the  result of fluctuations in experimental measurements.  It  is similarly stated in
 Ref. \cite{Erler:2004nh}  that at least a part of the problem in this case
  may be experimental.  The above appears to   indicate that $A_{FB}^{(0,b)}$
  is  on a somewhat less firm footing than  the other electroweak parameters.
The constraints on the $Z'$ of StLR are very similar to the
constraints  on the $Z'$ arising in StSM and we do not give a
separate detailed analysis of it here.


\section{Comparison of the Stueckelberg  $Z'$ and Classic $Z'$ Models }
\subsection{The Stueckelberg  $Z'$ and the CDDT Parametrization }
It is  instructive  to compare the Stueckelberg  $Z'$ model with
other $Z'$ models. For this purpose it is  convenient to use the
parametrization of the orthogonal matrix $R$   in terms of angles
 \cite{kn3}
\be{R}=\left(
\begin{array}{ccc}
 c_ {\psi }c_ {\phi} -s_ {\theta} s_ {\phi}s_ {\psi } & -s_ {\psi }c_ {\phi} -s_ {\theta} s_ {\phi} c_ {\psi } & -c_ {\theta} s_ {\phi}  \\
 c_ {\psi }s_{\phi} +s_ {\theta} c_ {\phi} s_ {\psi } & -s_ {\psi }s_ {\phi} +s_ {\theta} c_ {\phi} c_ {\psi } & c_ {\theta}c_ {\phi}  \\
 -c_ {\theta} s_ {\psi } & -c_ {\theta} c_ {\psi } & s_ {\theta}
\end{array}
\right), \label{Rmatrix} \ee
where
\be\tan (\phi )=\frac{M_2}{M_1}= \epsilon,\hspace{.5cm}\tan (\theta
)=\frac{g_Y}{g_2}\cos (\phi )=\tan \left(\theta _W)\right.\cos (\phi
),\ee
\be\tan (\psi )=\frac{ \tan (\theta ) \tan (\phi )M_W^2}{\cos
(\theta )\left(M_{Z^{{\prime }}}^2-M_W^2\left(1+ \tan ^2(\theta
)\right)\right)}.\ee
The SM limit, again, corresponds to  $ \epsilon \to 0$  which
implies $\tan(\phi) ,  ~\tan(\psi) \to 0$ and
 $\theta \to \theta_W $.
 Using Eq. (\ref{Rmatrix}) we may write the photon field $A_{\mu}^{\gamma}$ in the form
\be A^{\gamma}_{\mu}  = -c_ {\theta} s_ {\phi}C_{\mu} +c_ {\theta}
c_ {\phi }B_{\mu} + s_ {\theta}A^{3}_{\mu}, \ee
which shows  that the photon field contains a component outside of
the set $(B_{\mu},~ A^{3}_{\mu})$ while in the conventional $Z-Z'$
models the photon field  is just a linear  combination of the fields
$(B_{\mu},~ A^{3}_{\mu})$. This is what sets  the StSM model apart
from the conventional models.  To carry out the  comparison with the
$Z-Z'$ models a bit further we might try to  mimic the $Z-Z'$ models
by introducing  ``\ {\it  rotated fields\ }'' $\tilde B^{\mu}_Y$ and
$\tilde  C^{\mu}$ \beqn
\tilde B^{\mu}_Y&=& B^{\mu}\cos\phi -C^{\mu} \sin\phi\nonumber\\
\tilde C^{\mu}&=& B^{\mu}\sin\phi+C^{\mu} \cos\phi \label{azzp2},
\eeqn where the  rotation depends only on $\epsilon$.  In terms  of
new  variables  the physical vector fields  in StSM are \beqn
A^{\mu}_{\gamma}&=& W^{3\mu} \sin\theta +\tilde B^{\mu}_Y \cos\theta\nonumber\\
Z^{\mu} &=&( W^{3\mu} \cos\theta-\tilde B^{\mu}_Y\sin\theta)\cos\psi+\tilde C^{\mu}\sin\psi\nonumber\\
Z'^{\mu} &=&\tilde C^{\mu} \cos\psi -(W^{3\mu} \cos\theta -\tilde
B^{\mu}_Y \sin\theta)\sin\psi, \label{azzp1} \eeqn where $W^{3\mu}
\equiv A^{3\mu}$. The mass terms for a generic $Z-Z'$ mixing model
with the  gauge group $SU(2)_L\times U(1)_Y\times U(1)_Z$  are
typically given by \cite{Carena:2004xs}
\begin{equation}\frac{v_{H_1}^2}{8} (gW^{3\mu}-g_YB^{\mu} -z_{H_1}g_Z B_Z^{\mu})^2  +
\frac{v_{H_2}^2}{8} (gW^{3\mu}-g_YB^{\mu} -z_{H_2}g_Z B_Z^{\mu})^2
+\frac{v_{\phi}^2}{8} (z_{\phi} g_Z B _Z^{\mu})^2 \label{azzp3}
\end{equation}
where $g_Z$ is the $U(1)_Z$ gauge coupling constant  and $B_Z^{\mu}$
is used to denote the $U(1)_Z$ gauge field. Here
 the eigenvectors for the photon, $Z$ and $Z'$ are as follows
\beqn
A^{\mu}_{\gamma}&=& W^{3\mu} \sin\theta_W + B^{\mu} \cos\theta_W\nonumber\\
Z^{\mu} &=& (W^{3\mu} \cos\theta_W -B^{\mu} \sin\theta_W )+ \epsilon_Z B^{\mu}_Z \nonumber\\
Z'^{\mu} &=& B^{\mu}_Z  -  \epsilon_Z(W^{3\mu} \cos\theta_W -
B^{\mu} \sin\theta_W) \label{azzp4} \eeqn where
\begin{equation}
\epsilon_Z= \frac{\delta M^2_{ZZ'}} { M_{Z'}^2-M_Z^2}, \label{azzp5}
\end{equation}
and where $M_Z$, $M_{Z'}$ and $\delta M^2_{ZZ'}$ are given by \beqn
M_Z^2&=& \frac{g^2(v_{H_1}^2 +v_{H_2}^2)  }{4\cos^2\theta_W}[1+O(\epsilon^2_Z)]\nonumber\\
M_{Z'}^2&=& \frac{g^2_Z}{4}   (z_{H_1}^2v_{H_1}^2+z_{H_2}^2v_{H_2}^2+   z_{\phi}^2 v_{\phi}^2)[1+O(\epsilon^2_Z)]\nonumber\\
\delta M^2_{ZZ'} &=&-\frac{gg_Z}{4\cos\theta_W} (z_{H_1}v_{H_1}^2 +
z_{H_2}v_{H_2}^2  ) \label{azzp6} .\eeqn Using the rotated  fields
one  finds that  there is some similarity between the expressions
for the physical fields  in  Eq.
 (\ref{azzp1})  and in  Eq. (\ref{azzp4}).  However, this similarity
is superficial and a closer  scrutiny of the mass matrices reveals
that there is no limiting procedure  connecting the sets  of
expressions. Of course  this  should be  rather obvious since the
symmetry breaking in the $Z-Z'$ models  arises only from the  Higgs
sector while in StSM such a breaking arises  both from the Higgs
sector and from the Stueckelberg  sector. Further, in $Z-Z'$
analyses $\epsilon_Z$ is severely constrained by LEP data (
$|\epsilon_Z| \lesssim 10^{-3}$) and is either neglected
\cite{Carena:2004xs, Kang:2004bz} in the diagonalizaton procedure or
the case considered is $z_{H_2} =0$ with $\tan\beta = v_{H_2}/
v_{H_1} \gtrsim 10$ . In either case, these extensions do not allow
for narrow resonances of MeV size widths.
 The mass matrix  given in Eq. (\ref{massmatrix})  is also valid for
the minimal Stueckelberg Supersymmetric   Standard Model [StMSSM]
\cite{kn2}. Some of the experimental implications of StSM and of
StMSSM particularly
 with regard to the $e^+e^-$ colliders were
investigated in Ref. \cite{kn3}. However, the implications at
hadron colliders and specifically at the LHC were not discussed and
this is the main topic of discussion in this  paper. In summary the
Stueckelberg  extended  models form a new class outside the
framework of the usual $Z-Z'$ mixing models given generically by
Eqs. (\ref{azzp3}-\ref{azzp6}) and there  is no limiting procedure
connecting these models with the StSM.

 \section{LHC Observables and Constraints on the StSM Parameter Space}
 \subsection{Drell-Yan Cross Section for $pp\to Z'\to l^+l^-$}
Next we discuss the production of the narrow $Z'$  by the Drell-Yan
process at  the LHC. For the hadronic process $A +B \to  V +X$,  and
the partonic subprocess $q  \bar{q}\to V \to l^+l^- $, the dilepton
doubly differential cross section to next to leading order (NLO)  is
given by
\beqn \frac{{d^2\sigma _{{{  AB}}} }}{{dM^2 dz}} = K
\frac{1}{s}\sum\limits_q {\left[ {\frac{d \sigma_{q
\bar{q}}^{SM}}{{dz}}}  + {\frac{d \sigma_{q \bar{q}}^{St-SM}}{{dz}}}
+ {\frac{d \sigma_{q \bar{q}}^{St}}{{dz}}} \right]} {\mathcal
W}_{\left\{ {{ AB}(q\bar q)} \right\}} (s,M^2 ). \label{general}
\eeqn
\beqn\mathcal{W}_{\left\{{ AB}\left(q\bar{q}\right)\right\}}(\tau
)=\int _0^1\int _0^1 d x d y \delta (\tau - x
y)\mathcal{P}_{\left\{{ AB}\left(q\bar{q}\right)\right\}} (x,y),
 \nonumber\\
\mathcal{P}_{\left\{{ AB}\left(q\bar{q}\right)\right\}}
(x,y)=f_{q,A}(x)f_{\bar{q},B}(y)+f_{\bar{q},A}(x)f_{q,B}(y).
\label{NLODY}\eeqn
Here the dimensionless variable $\tau  = M^2 /s $ relates the
invariant mass $M$ of the final state lepton  pair to the center of
mass energy $\sqrt s$ of the colliding hadrons and
$z=\cos{\theta^*}$, where $\theta^* $ is the angle between an
initial state parton and the final state lepton in the C-M frame of
the lepton anti-lepton pair. The term $d\sigma^{SM}/dz$ is the
Standard Model contribution, $d\sigma^{St}/dz$ is the contribution
from the Stueckelberg sector, and $d\sigma^{St-SM}/dz$ is the
interference term between the Standard Model and the Stueckelberg
sectors. The parton distribution functions (PDFs) which we denote by
$f_{q,A}(x)$ give the probability that a parton of type $q$ has a
fracton $x$ of the total hadron four momentum. The dependence  of
$f_{q,A}(x)$ on the mass factorization scale $ Q=M $ is implicit.
 For
 the LHC $A = B = p$, and one must note that quite generally that $f_{q,A}=f_{\bar{q},\bar{A}}$ and $f_{\bar{q},A}= f_{q,\bar{A}}$. The Drell-Yan $K$ factor is as discussed in detail in  Refs. \cite{Hamberg:1990np,Carena:2004xs,leike,Baur:2001ze,Kumar:2006id}.
The invariant dilepton differential cross section is at NLO \beqn
\frac{d \sigma _{AB}}{d M}= K \frac{2M}{s}  \sum _{q  }
\sigma_{q\bar{q}}\left(M^2\right)\mathcal{W}_{\left\{{
AB}\left(q\bar{q}\right)\right\}}(\tau), \eeqn
where the partonic cross section, ${\sigma}_{q\bar{q}}$, is defined
by integrating the term in square brackets of Eq. (\ref{general})
over the variable $z$ and is computed in Ref.  \cite{kn3}.  While
$d\sigma/dM$ is sensitive to the interference term, the integral
over $dM$ is not. Thus for the computation of $d\sigma/dz$ one may
just use the $Z'$ pole contribution in Eq. (\ref{general}). Using
the analysis of Ref.\cite{kn3} for the partonic process $q\bar q\to
l^+l^-$ one finds that for  $pp$ collisions  the
integration of the third term of Eq. (\ref{general}) over $M^2$
yields the angular distribution for the StSM $Z'$ model
\beqn \frac{{d\sigma _{{{ AB}}} }}{{dz}}
=\frac{K}{s}\sum\limits_q{\mathcal W}_{\left\{ {{ AB}(q\bar q)}
\right\}} (s,M^2_{Z'} ) \frac{G_F^2 M_Z^4 M_{Z'}}{48 \Gamma
_{Z'}}\left[ (1 + z^2 )(a_e^{\prime 2}  + v_e^{\prime 2}
)(a_q^{\prime 2}  + v_q^{\prime 2} ) \right]. \label{Stz} \eeqn
A further  integration over $z$ gives the production cross section
for the Stueckelberg $Z'$ gauge boson \beqn \sigma_{AB}  \cdot Br(Z'
\to l^ +  l^ -  )  = K \frac{\pi }{{6s}}\sum\limits_q {C_q }
{\mathcal W}_{\left\{ {{ AB}(q\bar q)} \right\}}(s,M_{Z'}^2), \eeqn
where  dimensionless $C_q$ are  given by
\begin{equation}
 C_q = 2g_M^2 Br(Z' \to l^ +  l^ -
)(a'^{ 2}_q  + v'^{ 2}_q ), \hspace{0.5cm} q=u,d
\end{equation}
and where $g_M^2  = \sqrt 2 G_F M_Z^2$.
  The   $C_u-C_d$ parameterization is
  as defined in Ref. \cite{Carena:2004xs} \footnote{We  note that the analysis of Ref. \cite{Carena:2004xs}  absorbs  a factor of 8 in their PDFs contained within the function, defined as $W_{Z'}$} and allows one to use experimental limits set on the dilepton final state production cross section without making reference to the PDFs;    the couplings of a particular model are needed only,  if the experimental limits are known. In fact, such a paramerization is perhaps the first step in solving the potential "LHC inverse problem" \cite{Arkani-Hamed:2005px} for the case of the $Z'$  as one can directly map between the signature space and the parameter space in a very simple way.
The relation between $C_u$ and $C_d$ is
        \be
        \frac{C_u}{C_d}=\frac{(v'^{ 2}_u  + a'^{ 2}_u )}{(v'^{ 2}_d  + a'^{ 2}_d )}
        \sim \frac{\mbox{Br}(Z' \rightarrow u\bar u)}{\mbox{Br}(Z' \rightarrow d\bar d)}. \label{eq:cucdstsm}
        \ee
 Although $C_{(u,d)}$ are functions of $\epsilon$ for the StSM, the ratio is in fact independant of  $\epsilon $.   The formulas given in this section
 are  also valid for the case of the StLR via transcribing the couplings as laid out in  Eq.
 (\ref{StLR-coup}).
 \subsection{Constraints  on the StSM Parameter  Space from the CDF and D\O\ Data}
\begin{figure}[htb]
  \begin{center}
    \epsfig{figure=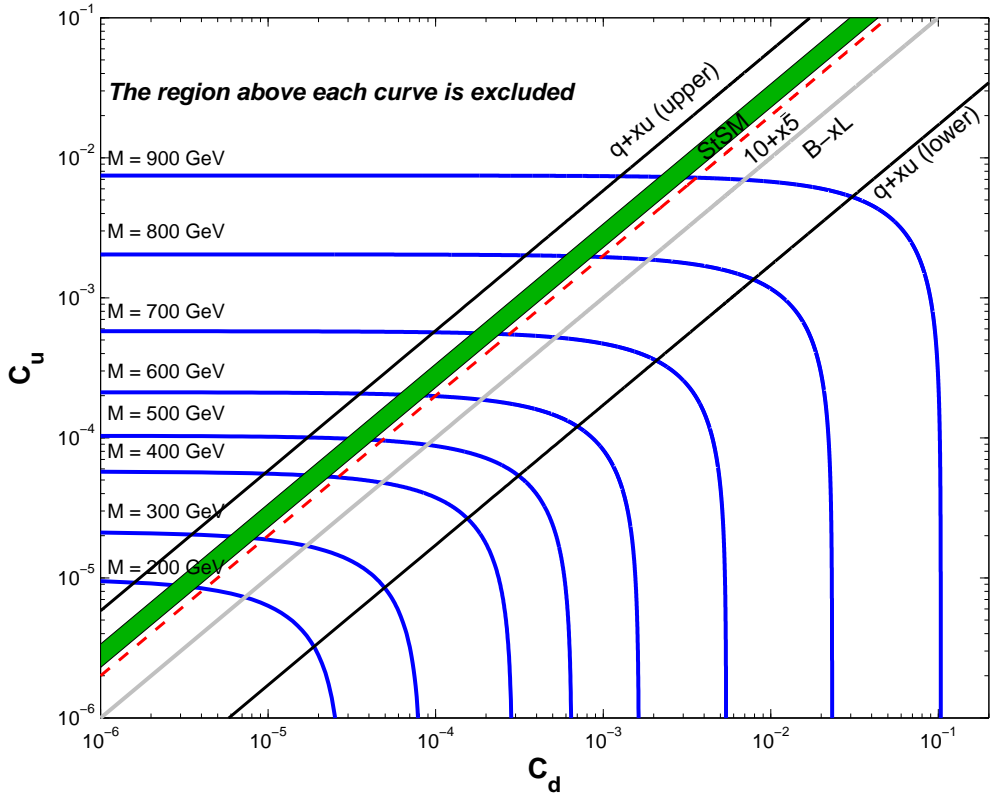,width=.9\textwidth}
\caption{Excluded regions in the $C_u-C_d$ plane from the current
95\% C.L. limit for $\sigma \cdot \mbox{Br}(Z'\rightarrow
l^{+}l^{-})$ given in \cite{cdf-z} at 819 ${\rm pb}^{-1}$  for
different $Z'$ masses, labeled as $M$ in the figure. The shaded
green band  is the region where the StSM model lies and
 where  $2.49 C_d < C_u < 3.37 C_d$. The light straight line corresponds to  $C_u$ and
$C_d$ in the $B-xL$ model where $C_u=C_d$ (see
\cite{Carena:2004xs}). The area between the two black straight lines
is the region where the $q+xu$ model lies and where $(3-2\sqrt{2})
C_d < C_u < (3+2\sqrt{2}) C_d$. The $10+x\bar{5}$ model is
constrained below the dashed red line which corresponds $C_u = 2
C_d$.} \label{cucd}

  \end{center}
\end{figure}

As  discussed  above the $C_u$-$C_d$ parametrization
\cite{Carena:2004xs}
 provides a useful technique to explore the limits on  new  physics and
  allows one to distinguish among various classes of models.
For instance, in the $C_u-C_d$ plane the  $C_u$ and $C_d$ predicted
in the StSM lie inside a  band.  The band structure for  StSM arises
since the ratio $C_u/C_d$ as given by  Eq. (\ref{eq:cucdstsm})  lies
in the range 2.49 $\sim$  3.37 for  $M_{Z'}$ lying in the range
$200 \sim 900$
  GeV.  Similarly,  the  $C_u$ and $C_d$ predicted  in   the   $q+xu$ model \cite{Carena:2004xs}
also lie in a band,  while the $C_u$ and $C_d$  for  the $B-xL$
model \cite{Carena:2004xs}   live on a line. In Fig. (\ref{cucd}) we
give  a  numerical evaluation of the $C_u$ and $C_d$ using the most
recent  CDF data of  819 ${\rm pb}^{-1}$ in the dilepton channel
\cite{cdf-z}.
 The $C_u$-$C_d$ exclusion plots of  Fig. (\ref{cucd})
  can be used to constrain $\epsilon$ for a given $M_{Z'}$.
These  constraints  are consistent with the constraints derived
using a smaller data sample  of approxomately 275 ${\rm pb}^{-1}$
which, however, uses  the  more sensitive D\O\ mode
\cite{Abazov:2005pi}. In addition to the above   one also has
constraints on the parameter space from the non-observation of the
$Z'$ from the CDF and D\O\ data
\cite{cdf-z,Abazov:2005pi,cdfdata,Abulencia:2006iv}.
 These constraints were shown to
limit values of  $(\epsilon,M_{Z'})$ in \cite{Feldman:2006ce}, while
still allowing for the possibility of a narrow StSM $Z'$   which
could even lie  relatively close to   the $Z$-pole.

\section{Discovery Reach of LHC for StSM $Z'$ Boson}
\subsection{$\sigma \cdot Br(Z'\to l^+l^-)$ at the LHC}
\begin{figure}[htb]
  \begin{center}
  \epsfig{figure=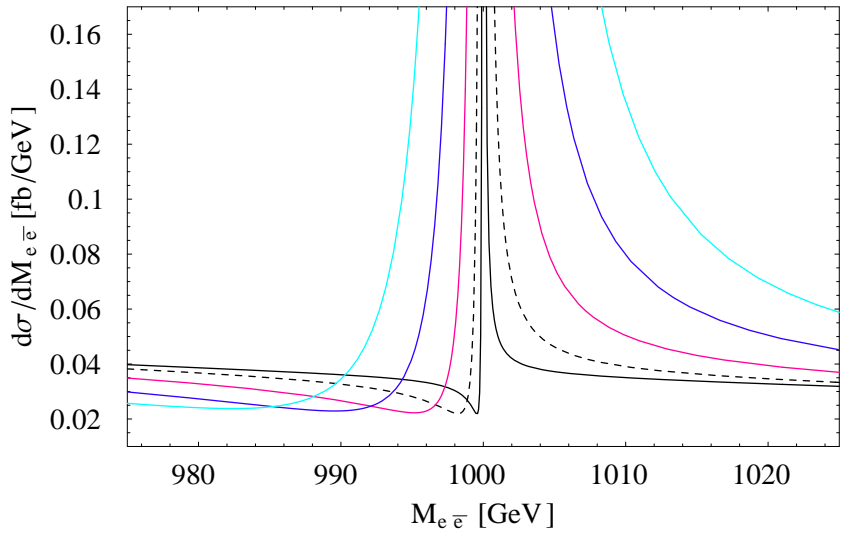,width=.8\textwidth}
\caption{\label{Rainbow} The invariant dilepton differential cross
section, $d\sigma(pp \to Z'\to l^+\l^-)/dM$ as a  function of the
dilepton invariant  mass for various $\epsilon$ values. The plot
exhibits the narrow widths at the $Z'$ pole. The dashed curve
corresponds to $\epsilon =.06$. The shapes of these curves
illustrate the exceptionally narrow resonance  widths of the StSM
$Z'$ with
 distinct distributions. }

  \end{center}
\end{figure}

Next  we give an analysis for the  exploration of  the $Z'$  boson
at  the LHC. Before  proceeding further it is instructive  to
examine the shape of the $d\sigma/dM$ as a function of the invariant
mass $M$. This is exhibited in Fig. (\ref{Rainbow}) where the plots
are given for an array  of values of $\epsilon$ (ranging over the
set  $ \{.03, .06, .1, .15, .2\}$ where  the larger  values of
$\epsilon$ are taken only for illustrative purposes)  for the case
when $M_1=1$ TeV. One can appreciate the narrowness of the $Z'$ pole
from these plots. This type  of shape  and  width is strikngly
different from the  ones encountered  in the conventional $Z'$
models \cite{leike} and also in Kaluza-Klein excitations of the $Z$
boson in  large radius extra  dimension models
\cite{Antoniadis:1999bq,Nath:1999mw}.

The quantity that will be measured  experimentally at the LHC  is
$\sigma_{pp} \cdot Br(X \to l^+l^-) \equiv \sigma \cdot Br(X \to
l^+l^-)$ in the process $pp\to X\to l^+l^-$
 where $X$ is a neutral resonant state produced in $pp$ collisions which can decay into a lepton pair.  Here
 we  give  a theoretical analysis of this quantity for the case when $X=Z'$,  and in  the next  section we will consider  the case
 when $X=G$,  the spin 2 graviton of a  warped geometry.
In the analysis of $\sigma \cdot Br(Z'\to l^+l^-)$ we will discuss
two regions: a low mass region with the dilepton invariant mass
$M_{l \bar l}$ up to 800 GeV and a high mass region with $M_{l \bar
l}$ extending from 800 GeV up to the maximum relevant mass  reach of
the LHC. The reason for this ordering is as follows: the region with
$M_{l \bar l}$ up to 800 GeV has already begun to be explored at the
Tevatron using up to about $1~\rm fb^{-1}$ of data, and the CDF and
D\O\ data puts constraints on $\epsilon$ as a function of the
dilepton invariant mass. Thus in the analysis of the low mass $M_{l
\bar l}$ region at the LHC we can incorporate these  constraints.
However, one has no direct constraints in the dilepton invariant
mass region above 800 GeV, which explains the separate analyses of
$\sigma \cdot Br(Z'\to l^+l^-)$ for the low and high mass regions.

\begin{figure}[htb]
  \begin{center}
    \epsfig{figure=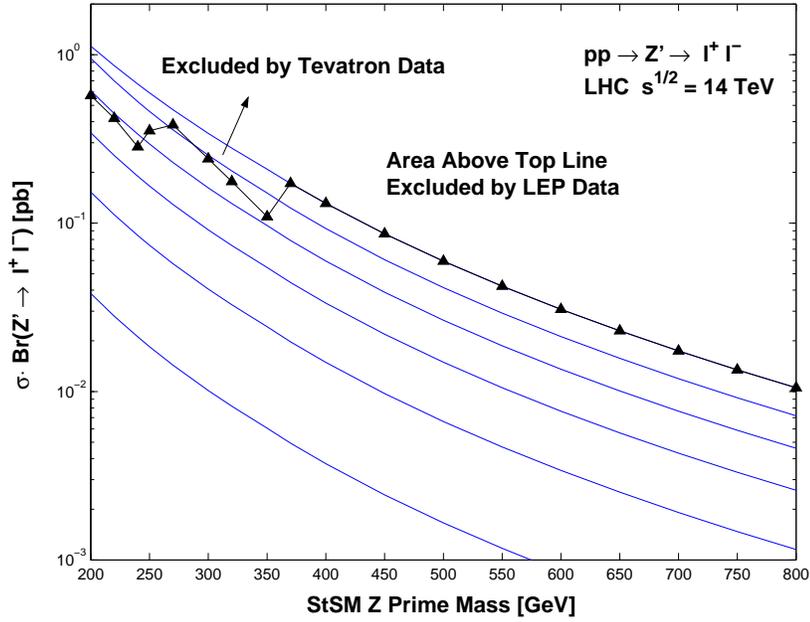,width=.8\textwidth}
\caption{The production cross section $ \sigma \cdot Br(Z' \to l^+
l^-) $ [pb]   in the StSM at the LHC  in the low mass region with
the inclusion of the LEP and Tevatron constraints. The curves in
descending order  correspond  to values of $\epsilon$ from .06 to
.01 in steps of .01. }\label{lowmassdata}
  \end{center}
\end{figure}

\begin{figure}[htb]
  \begin{center}
\includegraphics[width=13.5cm,height=11cm]{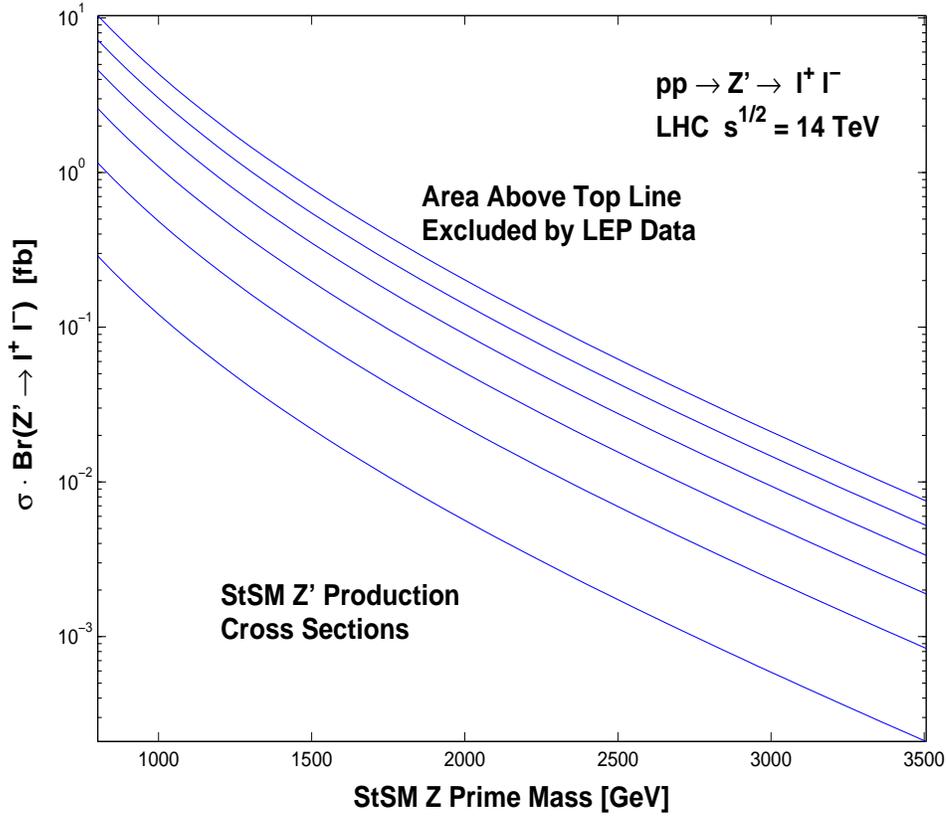}
\caption{The production cross section
 $ \sigma \cdot Br(Z' \to l^+ l^-) $ [fb]  in the StSM  at the LHC
in the  $Z'$  high mass region  up to $Z'$ mass of $\approx$ 3.5
TeV. The curves correspond to values of $\epsilon$ ranging from
$.06$ to $.01$ in descending order in steps of .01. The StSM
production cross sections sit several orders of magnitude below
those of other $Z'$ models. } \label{HighmassLHC}
  \end{center}
\end{figure}

We begin with an analysis of  $\sigma \cdot Br(Z'\to l^+l^-)$ in the
low mass region where we use the constraints on $ (\epsilon,M_{Z'})$
as obtained  in Ref.  \cite{Feldman:2006ce} using the cross section
limits from \cite{Abazov:2005pi}.  The results are displayed in Fig.
(\ref{lowmassdata}).  As expected  one  finds that the current data
on $\sigma \cdot Br(Z'\to l^+l^-)$ constrains only the  mass region
of $Z'$ for values $M_{Z'}\lesssim 350$ GeV. We note that for
$\epsilon$ as high as $ \approx .04 $  one may have an StSM $Z'$  as
low as 175 GeV, while with a $Z'$ mass of 250 GeV, $\epsilon$ may be
as high as $\approx .035$ within the current experimental limits.
Next we discuss the high mass region for the StSM  $Z'$. As
discussed above the high mass region of StSM $Z'$
 remains unconstrained by the CDF and D\O\ data, and thus in this region only the LEP electroweak constraints apply.
 The analysis  of  Fig. (\ref{HighmassLHC})  gives a  plot of $ \sigma \cdot Br(Z'\to l^+l^-)$
 as a function of $M_{Z'}$   in the high mass region
 for values of $\epsilon$
ranging from  $.01$ to  $.06$  in ascending order in steps of $.01$.
From  Fig. (\ref{HighmassLHC}) and from the analysis  of Refs.
\cite{Kang:2004bz,Dittmar:2003ir} for other $Z'$ models one infers
that the production cross section
 for StSM $Z'$ lies  orders of magnitude below those for the  $Z'$ production in  E6 models and other $Z'$
 models. The size of   $ \sigma \cdot Br(Z'\to l^+l^-)$ thus provides a clear
 signature which differentiates the StSM $Z'$ model from other $Z'$ models.

 \subsection{Signal to Background Ratio}
\begin{figure}[htb]
  \begin{center}
\includegraphics[width=14cm,height=11.5cm]{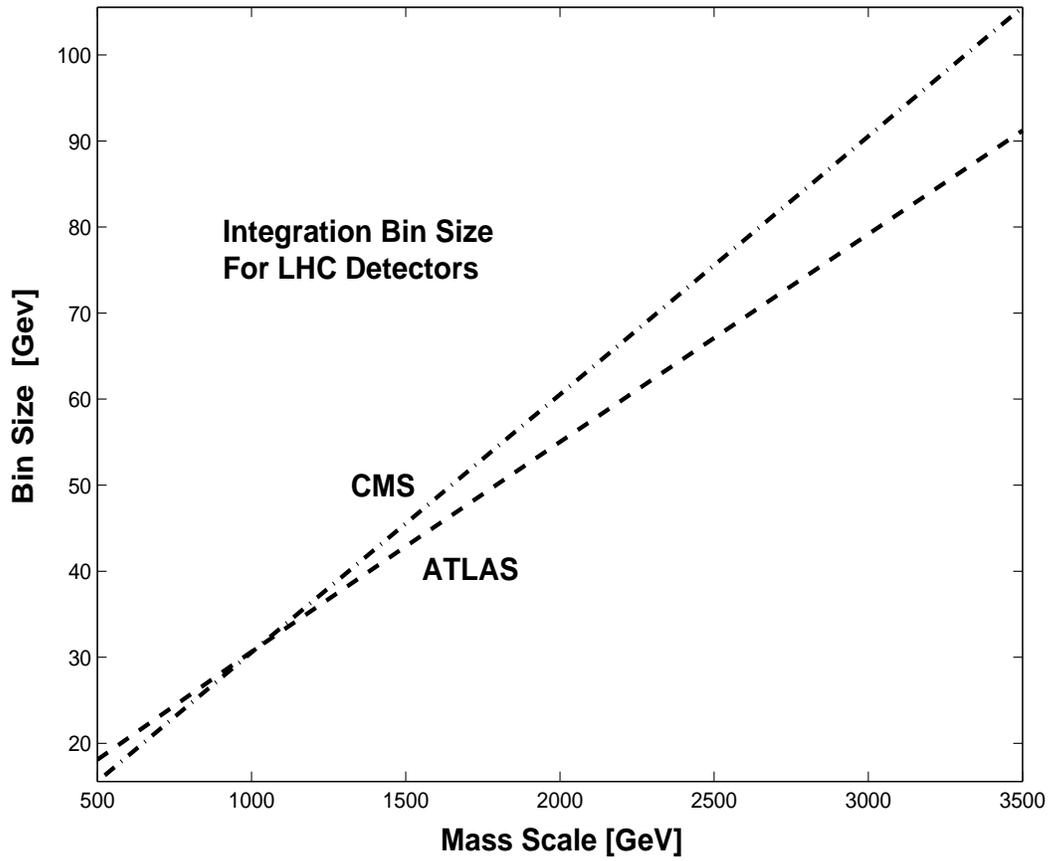}
\caption{ A plot of the mass window or bin size as a function of the
mass scale for the ATLAS and CMS detectors.} \label{binsize}
  \end{center}
\end{figure}

\begin{figure}[htb]
  \begin{center}
\includegraphics[width=12.5cm,height=11.5cm]{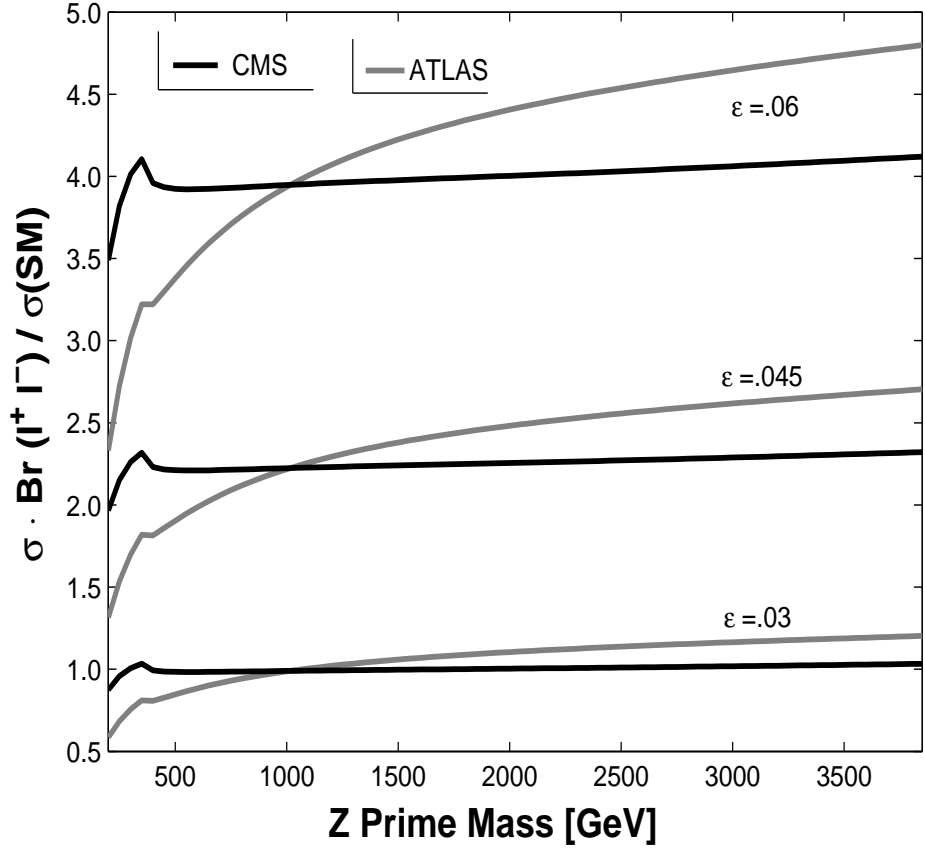}
\caption{A plot of the ratio $\sigma \cdot Br(Z'\to
l^+l^-)_{StSM}/\sigma_{SM}(Z, \gamma\to l^+l^-)$ including the
$\gamma - Z$ interference term in the SM as a function of the $Z'$
mass for the ATLAS and CMS detectors assuming the bin sizes as in
Fig.(\ref{binsize}) for values of $\epsilon$ in the range .03-.06.
The signal to background ratio is larger for the CMS detector at low
mass scales while it is larger for the ATLAS detector at  large mass
scales   with a cross over occuring at around 1 TeV.}
 \label{signoise}
  \end{center}
\end{figure}

 The dilepton channel will be
analyzed at the LHC in the ATLAS \cite{Atlas}  and CMS \cite{GOA}
detectors, and as is  discussed below, both detectors have  the
ability
 to probe the  narrow StSM $Z'$ boson.
Experimentally, the discovery of a narrow resonance depends to a
significant degree
 on the  bin size  for data collection with the chance of detection increasing with a decreasing bin size. This is so because the integral
over the bin is  effectively  independent of the bin size for the
signal (assuming the narrow resonance falls within the bin).
However, this integral is essentially linearly dependent on the bin
size for the SM background.
  In the analysis of the SM background  we have  included  the $Z, \gamma$, and $ \gamma-Z$
 interference terms in the Drell-Yan analysis, but  have not included the backgrounds from other
 sources such as from $t\bar t, b\bar b,  WW, WZ, ZZ$ etc.  However, these backgrounds  are known to be
 at best a few percent of the Drell-Yan background \cite{Cousins}.
 Regarding the bin size,  it  depends on the energy resolution
$\sigma_E/E$ of the calorimeter.  For an electromagnetic calorimeter
the energy resolution is typically parameterized  by $\sigma_E/E=
a/\sqrt E  \oplus  b \oplus c/E$  where  addition in quadrature is
implied\cite{pdg}. The term proportional to  $1/\sqrt E$ is  the so
called stochastic  term  and arises from statistic  related
fluctuations. The term $b$ is due to detector  non-uniformity and
calibration errors, and the term $c$ is due  mostly  to noise.
For the ATLAS detector (liquid Ar/Pb) the energy resolution is
 parameterized by \cite{pdg} $\sigma_E= 10\%/\sqrt E \oplus .4\% \oplus .3/E$ and for the CMS detector ($\rm PbWO_4$)  it  is parameterized by
 $\sigma_E= 3\%/\sqrt E \oplus .5\% \oplus .2/E$ where $E$ is in units  of GeV. From the above we find  the  following relations for
 the bin size B (taken to be 6$\sigma_E$)  at the mass scale  $M$ ($M$ is measured in units of TeV)
  \beqn
 {\rm B}_{\rm ATLAS} = 24 (.625 M +  M^2  +.0056)^{1/2} {\rm GeV}\nonumber\\
 {\rm B}_{\rm CMS}  = 30 (.036 M + M^2 + .0016)^{1/2} {\rm GeV}.
 \label{twobins}
       \eeqn
For $M > 3$ TeV, the  $M^2$ term dominates  in Eq.(\ref{twobins})
and the  bin size goes linearly in $M$, so ${\rm B}_{\rm ATLAS}\sim
24 M$ GeV and ${\rm B}_{\rm CMS}\sim 30 M$ GeV for large $M$. A plot
of bin sizes as  a function of the mass scale  is  given  in
Fig.(\ref{binsize}) for the two  LHC detectors. One  finds  that  at
low  mass  scales the CMS has a  somewhat better  energy resolution
and thus a  somewhat smaller  bin sizes and  at  large  mass  scales
ATLAS  has  a somewhat better energy resolution and thus a somewhat
smaller bin size with a cross  over  at $M \sim 1$ TeV.   However,
on the whole the energy resolution and the  bin size  of the two
detectors  are comparable  within about 10\%. For the StSM $Z'$ the
analysis of Fig. (\ref{signoise}) shows that the signal to
background  is greater than unity in significant parts of the
parameter space, and in some cases greater than 4, thus illustrating
that the LHC has the ability  to detect a strong signal for a StSM
$Z'$.


\subsection{How  Large  a $Z'$ Mass  and How  Narrow a $Z'$ Width  Can LHC Probe? }
\begin{figure}[htb]
  \begin{center}
  \epsfig{figure=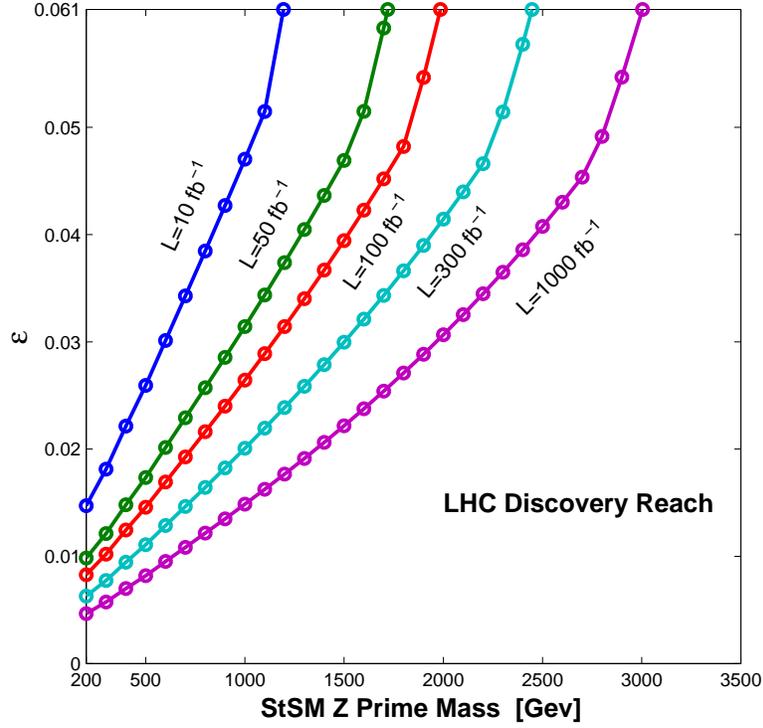,width=.7\textwidth}
 \epsfig{figure=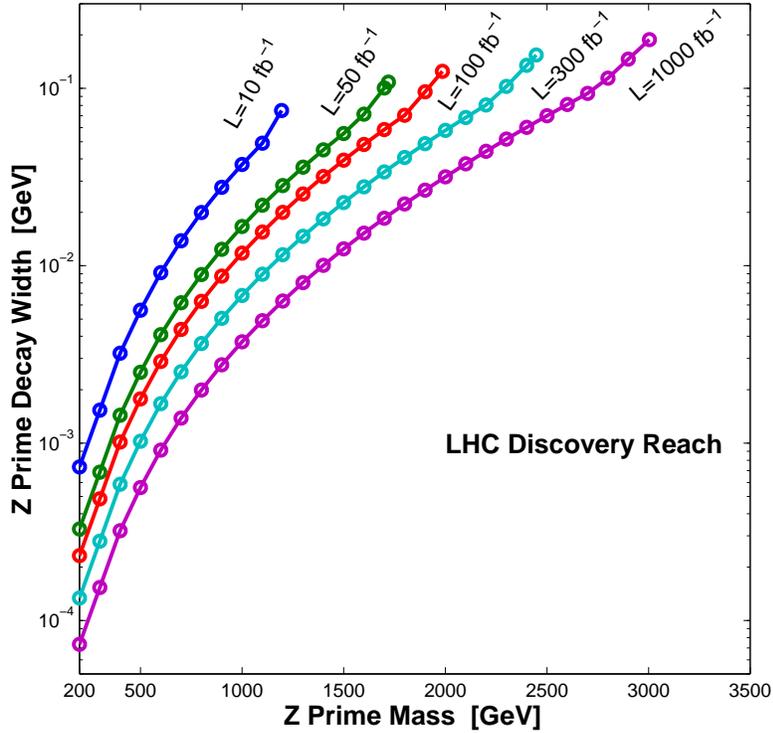,width=.7\textwidth}
\caption{ A plot of the discovery limits of $Z'$ in StSM with the
discovery limit defined by
 $5\sqrt {N_{SM}}$ or by 10 events, whichever is larger. The inflections, or kinks, in the plots
 are precisely the points of transitions between the two criteria. Regions to the left and above each curve can be probed by the LHC at a given luminosity.  The top point on each curve corresponds to $\epsilon =.061$.  The analysis is  done for the  ATLAS detector but similar results hold  for
 the CMS detector.}
\label{searchreachepslionandwidth2}
  \end{center}
\end{figure}

\begin{figure}[htb]
  \begin{center}
    \epsfig{figure=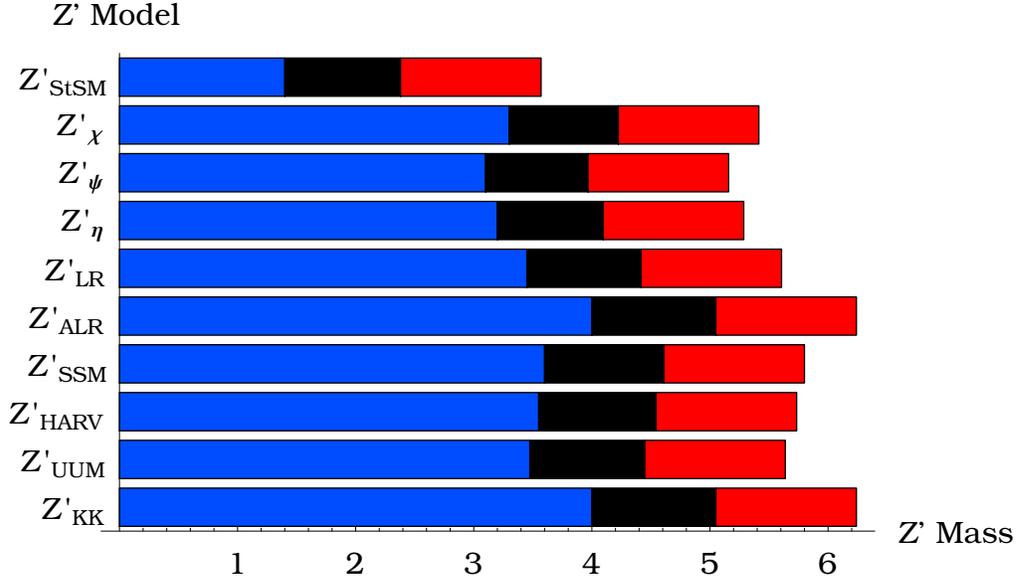,width=1\textwidth}
\caption{\label{bars} The discovery reach for   $Z'$ in StSM
(without detector cuts) and several other $Z'$  models at the LHC.
The length of the bars indicate integrated luminosities of 10
${\rm{fb}}^{-1}$ (blue), 100 ${\rm{fb}}^{-1}$ (black), and 1000
${\rm{fb}}^{-1}$ (red) using 10 events as  the criterion for
discovery \cite{Godfrey:2002tn,Cvetic:1995zs}.
 The analysis  indicates that the $Z'$ of StSM  can be probed up to $\approx$  3.5 TeV at the
LHC with 1000 ${\rm{fb}}^{-1}$ of integrated luminosity. With
inclusion of detector cuts the discovery reach of the LHC for the
StSM $Z'$ comes down to about 3 TeV.}
  \end{center}
\end{figure}

In Fig. (\ref{searchreachepslionandwidth2}) we give the discovery
reach for finding the StSM $Z'$ with various values of $\epsilon$ as
a function of $M_{Z'}$ for integrated luminosities in the range 10
${\rm{fb}}^{-1}$ to 1000 ${\rm{fb}}^{-1}$. The criterion used for
the discovery limit in the analysis given here is an assumption that
$5\sqrt {N_{SM}}$  events or  10 events, whichever is larger,
constitutes a signal
 where $N_{SM}$ is the SM background,  and we have scaled the  bin size with $M_{Z'}$ appropriate for
 the ATLAS detector with a conservative  lower limit  of 20 GeV below .5 TeV.
In this  part of the analysis  we have  assumed that detector
effects can lead to signal and background losses of 50 percent (see
Section (\ref{signature})). If better efficiency and acceptance cuts
are available, the discovery reach of the LHC for finding a $Z'$
  will be even higher than what  we  have displayed.
  With an assumption of efficiencies  as  stated  above,
one finds that with 100  ${\rm{fb}}^{-1}$ of integrated luminosity,
one can explore  a $Z'$ up to about 2 TeV with $\epsilon=0.06$, and
this limit can be pushed to $\approx$ 3 TeV with 1000
${\rm{fb}}^{-1}$ of integrated  luminosity. Further, one finds that
for 1000 ${\rm{fb}}^{-1}$ of integrated  luminosity,  one can
explore a $Z'$ up to about 2 TeV for $\epsilon$ as low as $\lesssim
0.02$. Also displayed in Fig. (\ref{searchreachepslionandwidth2})
are the discovery limits for different decay widths as a function of
the $Z'$ mass again for luminosities in the  range  10
${\rm{fb}}^{-1}$ and 1000  ${\rm{fb}}^{-1}$. Here one finds that the
LHC  can probe a 100 MeV $Z'$ up to about 2.75 TeV and a 10 MeV
width  up to a $Z'$ mass of about  1.5 TeV. A more detailed
exhibition of the capability of the LHC to probe the StSM $Z'$ model
is given in Fig. (\ref{submevwidth}). Here one finds that  the StSM
model  with a $Z'$ width even in the MeV and sub-MeV range will
produce a detectable signal in the dilepton channel in the Drell-Yan
process with luminosities accessible at the LHC. While the analysis
above is for the specific StSM model, the general features of this
analysis may hold for a wider class of models which support narrow
resonances.
In  Fig. (\ref{bars}) we give a comparison of the LHC's ability to
probe the narrow StSM $Z'$ relative to  other $Z'$ models
 \cite{Godfrey:2002tn,Cvetic:1995zs} to address the question
of how the StSM $Z'$ ``stacks up'' to these models. In order to make
the appropriate comparisons of the discovery limits for the StSM
with the other $Z$ prime models we do not impose detector cuts on
the StSM  $Z'$  limits  displayed in Fig. (\ref{bars}), since such
cuts were not imposed for the discovery limits of other $Z'$ models
shown in  Fig. (\ref{bars}). The analysis of Fig.
 (\ref{bars}) shows that the StSM $Z'$,  even with its exceptionally
narrow width,  may be probed on  scales comparable  with models that
have resonance widths of the order of  several GeV or higher.

\section{Comparison of Stueckelberg  $Z'$  with a Massive Graviton of Warped Geometry at the LHC }
As discussed above  one finds that the  Stueckelberg  $Z'$ boson  is
a very narrow resonance which sets it  apart from all other $Z'$
models.  However, there is another class of models,  i.e., models
based on warped geometry
 \cite{Randall:1999ee,Gogberashvili:1998vx} (labeled RS  models),  which can mimic
the   Stueckelberg $Z'$ in a  certain part of the parameter space as
far as  the narrowness of the resonance is  concerned. It was  shown
in the analysis of  Ref.  \cite{Feldman:2006ce}   that the signature
spaces for these  two models
 lie close  to each other  in certain regions of their respective parameter spaces,
but the models are  still distinguishable in the dilepton mass
region accessible at the Tevatron. Here we extend the analysis  of
their relative signatures  to the LHC energies. The geometry of RS
models  is  a slice of  $AdS_5$  described by the metric $ds^2$
=$exp(-2kr_c|\phi|)\eta_{\mu\nu}dx^{\mu}dx^{\nu} -r_c^2 d\phi^2$,
$0\leq \phi\leq \pi$, where $r_c$ is the radius of the extra
dimension and $k$ is the curvature of $AdS_5$, which is taken to be
the order of the Planck scale. We work in the regime where the SM
particles are confined to the TeV scale brane, while gravity is
propagating in the
 bulk  \cite{ Randall:1999ee,Davoudiasl:1999jd}.
  The effective scale that enters in the electroweak  region is the scale $\Lambda_{\pi}=\bar{M}_{Pl} exp(-kr_c\pi)$, and
 for reasons of naturalness it is typically constrained by the condition $\Lambda_{\pi}< 10$ TeV.
 Values of $k/{\bar{M}_{Pl}}$ over  a  wide  range $10^{-5}-.1$ have been considered in the literature \cite{Kisselev:2005bv}.
However,  the range  below  $.01 $ appears to be eliminated from the
electroweak constraints.
 In this analysis we consider  the lightest massive graviton mode .

%
%
\subsection{Drell-Yan Cross Sections via a Massive Graviton of Warped  Geometry}
We consider the process $pp \to G \to f \bar f$ for the first
massive graviton mode in the RS model. The partonic production cross
section for this mode receives  contributions both from quarks and
gluons, and is given by
\cite{Han:1998sg,Giudice:1998ck,Bijnens:2001gh,Dvergsnes:2002nc,Mathews:2004xp}
\be \frac{{d\sigma _{q\bar q}^G }}{{dz}} + \frac{{d\sigma _{gg}^G
}}{{dz}} = \frac{1}{2}\frac{{\kappa ^4 M^6 }}{{320\pi ^2 }}\left[
{\Delta _{q\bar q} (z) + \Delta _{gg} (z)} \right]\frac{1}{{(M^2  -
M_G^2 )^2  + M^2 \Gamma _G^2 }}. \ee The total decay width that
enters above is given by the sum of the partial widths which are
\cite{Han:1998sg,Allanach:2000nr,Bijnens:2001gh}
\beqn
&&\Gamma(G \to V \bar V) =\delta \frac{\kappa^2 M^3_G}{80 \pi}(1-4 {\delta}_V)^{1/2}\left(\frac{13}{12}+\frac{14}{3}{\delta}_V+4{\delta}^2_V\right) \theta(M_G - 2 M_V)\\
&&\Gamma(G \to f \bar f) = N^c_f \frac{\kappa^2 M^3_G}{320\pi}(1-4 {\delta}_f)^{3/2}(1+\frac{8}{3}{\delta}_f) \theta(M_G - 2 m_f) \\
&&\Gamma(G \to g g) =  \frac{\kappa^2 M^3_G}{20 \pi}\\
&&\Gamma(G \to \gamma \gamma) =  \frac{\kappa^2 M^3_G}{160 \pi}.
 \eeqn
Here  $\delta_f = m^2_f/M^2_G$, $\delta_V = M^2_V/M^2_G$, and
$\delta  = (1/2,1)$ for $( V = W,Z)$.
 For the first massive mode,   $\kappa$ is given by  \cite{Allanach:2000nr,Bijnens:2001gh,Dvergsnes:2002nc}
\be \kappa  =  \sqrt 2 \frac{{x_1 }}{{m_G }}\frac{k}{{\bar M_{Pl} }}
\ee where $x_1=3.8317$ is the first root of the Bessel function of
order 1, and $\bar{M}_{Pl}$ is the reduced Planck mass in four
dimensions ($\bar{M}_{Pl} = M_{Pl}/\sqrt{8 \pi}$).
The leading order angular dependance is given in terms of
\cite{Bijnens:2001gh,Dvergsnes:2002nc,Mathews:2004xp} \be \Delta
_{q\bar q} (z) = \frac{\pi }{{8N_c }}\frac{5}{8}(1 - 3z^2  + 4z^4 ),
\hspace{1cm} \Delta _{g g} (z) = \frac{\pi }{{2(N_c^2  -
1)}}\frac{5}{8}(1 - z^4 ).\label{angfuncs} \ee
In the narrow width approximation we have to NLO
\beqn
 \frac{d\sigma^G_{p p} }{dz}  &=&
K^G (M_G^2 )\frac{1}{{2s}}\frac{{\kappa ^4 M_G^6 }}{{320\pi ^2
}}\frac{\pi }{{M_G \Gamma _G }}\times
\\
&& \hspace{.5cm} \left[ {\sum\limits_q {\Delta _{q\bar q} (z)}
{\mathcal W}_{\left\{ {pp(q\bar q)} \right\}} (s,M_G^2 ) + {\Delta
_{gg} (z)} {\mathcal W}_{\left\{ {pp(gg)} \right\}} (s,M_G^2 )}
\right]\label{RSz} \nonumber \eeqn
where $ {\mathcal W}_{pp(q\bar q)}$ is defined  in Section 6 and $
{\mathcal W}_{pp(gg)}$ is defined by
\beqn\mathcal{W}_{\left\{{pp}\left(gg\right)\right\}}(\tau )=\int
_0^1\int _0^1 d x d y \delta (\tau - x y) f_{g,p}(x)f_{g,p}(y),
\label{NLODY2}\eeqn and the more strongly mass dependant RS $K$
factor ($K^G$)  is discussed in detail in  Refs.
\cite{Mathews:2004xp}.
The production cross section including  the quark and gluon
contributions is in the narrow width approximation given by
\beqn && \hspace{.5cm}
 \sigma  \cdot Br(G \to l^ +  l^ -  ) = K^G (M_G^2 )\frac{1}{s}\frac{{\kappa ^4 M_G^6 }}{{{\rm{15360}}}}\frac{1 }{{M_G \Gamma _G }}\sum\limits_q {{\mathcal W}_{\left\{ {pp(q\bar q)} \right\}} } (s,M_G^2 )
\\
&& \hspace{4 cm} + K^G (M_G^2 )\frac{1}{s}\frac{{\kappa ^4 M_G^6
}}{{{\rm{10240}}}}\frac{1}{{M_G \Gamma _G }}{\mathcal W}_{\left\{
{pp(gg)} \right\}}  (s,M_G^2 ).\nonumber \eeqn

\subsection{Signature Spaces of  StSM $Z'$ and of the Warped Geometry Graviton}\label{signature}

A relative comparison of the StSM and of the RS model is given in
Table (\ref{tab:table2}) where the decay width of  the Stueckelberg
$Z'$ boson for the case $\epsilon =0.06$ is given as  a function of
the $Z'$ mass in the range (1000-3000) GeV, and  the corresponding
$\sigma \cdot Br(G \to l^+l^-)$ is exhibited. Also shown are the
decay widths for an RS graviton in the same mass range for
$k/{\bar{M}_{Pl}} =0.01$.

 Quite remarkably, the spin 1 $Z'$ of the
StSM and the spin 2 massive graviton of the RS model have nearly
identical signatures in terms of the decay widths and the production
cross sections around a resonance mass of  2 TeV (with or without
out detector cuts). In Table (\ref{tab:table3}) we give an analysis
of the number  of events  that can be observed in the ATLAS
detector with 100 ${\rm{fb}}^{-1}$ of integrated  luminosity. One
finds that for high masses the number of events  that one expects to
see at the LHC for the StSM $Z'$, with $\epsilon =0.06$, are similar
to the number  of events one expects  for  the RS model for $k/\bar
M_{Pl} =0.01$.  For the case of the RS model, simulations conducted
by Ref. \cite{Allanach:2000nr} show that overall detector losses
range from (27-38) percent between (500-2200) GeV, and we have
extrapolated these cuts to the 3 TeV mass region.  For the case of
$Z'$, which has a different angular dependancy than the graviton due
to spin, we have assumed a uniform 50 percent loss of events at in
the range of $Z'$ mass investigated. This reduction factor is
consistent with the reduction factor used by Ref. \cite{SLT}, and is
similar to the reduction factor used by other groups
\cite{Rizzo:1996ce}. For the SM background, denoted as $N_{B} =
N_{SM}$, the same detector loss is assumed, and it can be seen in
Table (\ref{tab:table3}) that this simulation is in good agreement
with the analysis of Ref. \cite{Allanach:2000nr}. Of course  a
slightly more  realistic analysis  of the number of events that may
be observed requires simulating detector efficiencies more
accurately, which in turn requires the implementation of the StSM
couplings in event generation simulators
\cite{Traczyk:2002jh,Collard:2004ab,
Lemaire,Cousins,SLT}. \\

\begin{figure}[h]
\hspace*{-.2in} \centering
\includegraphics[width=16cm,height=12cm]{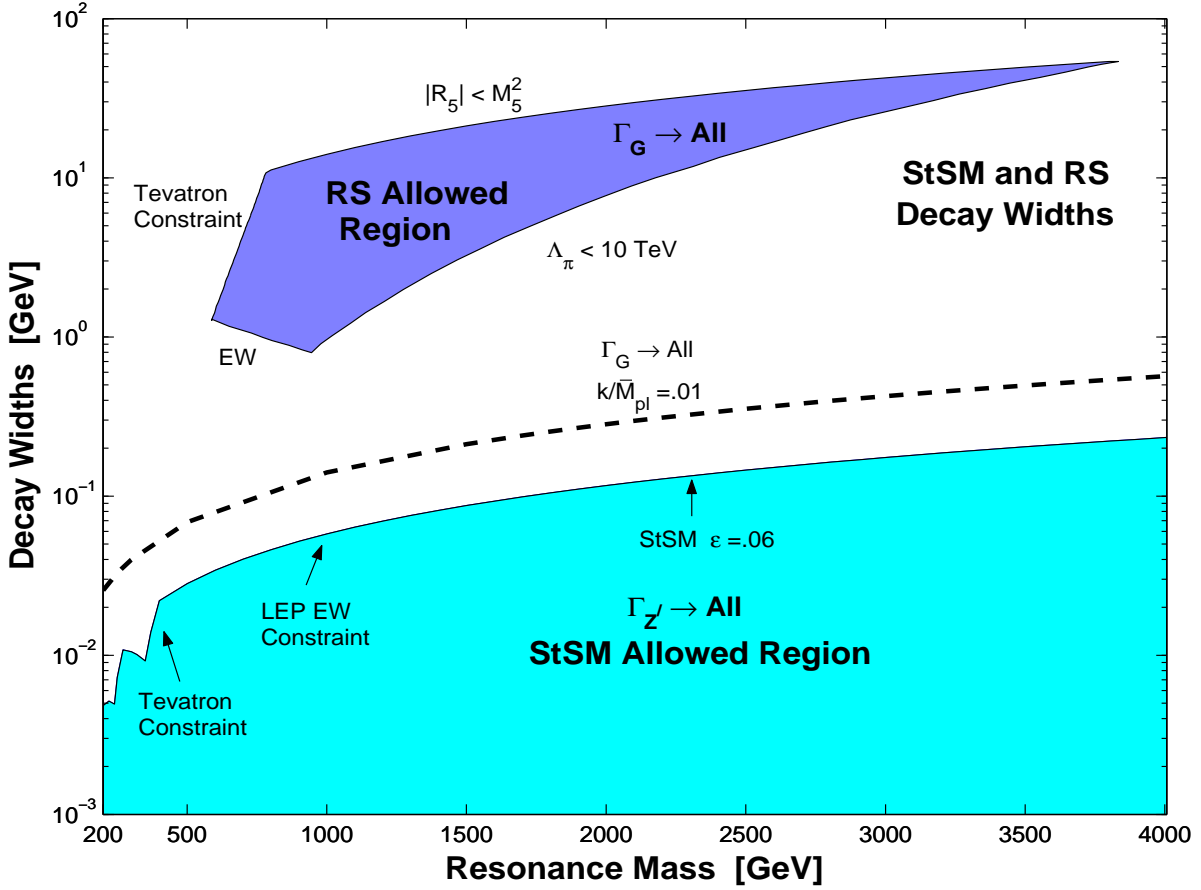}
\caption{ A comparison of the allowed region in resonance decay
width - resonance mass plane for the $Z'$ in the StSM and the first
graviton mode in the RS model.   The dashed line is for the RS case
with  $k/{\bar{M}}_{Pl} = .01$. The allowed  (shaded) regions are
 constructed by utilizing  the constrained parameter spaces of  StSM \cite{Feldman:2006ce} and
 the  RS model \cite{Davoudiasl:1999jd,Davoudiasl:2000wi,Abazov:2005pi}.}
\label{widths}
\end{figure}

\begin{figure}[h]
\hspace*{-.2in} \centering
\includegraphics[width=16cm,height=12cm]{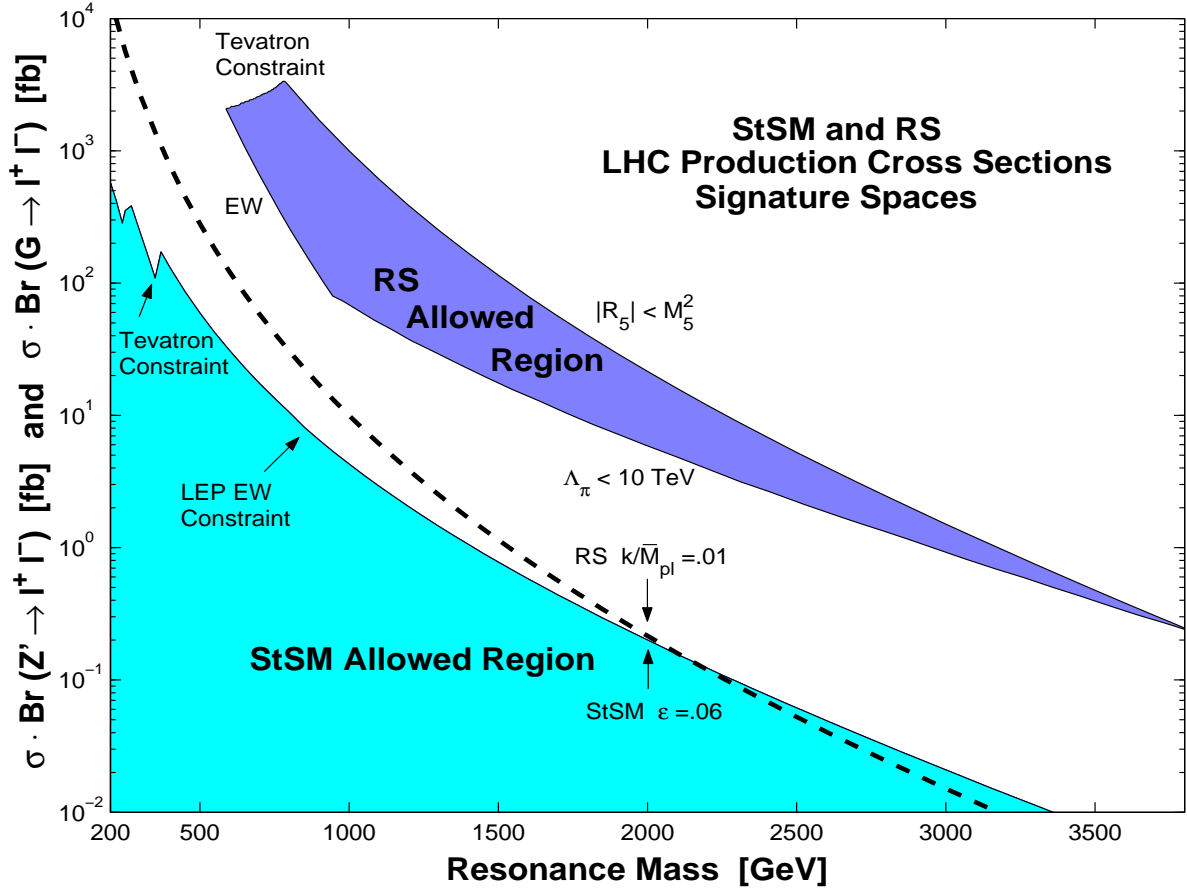}
\caption{A comparison of the LHC signature spaces in the dilepton
channel using $\sigma(pp\to Z'\to l^+l^-)$  for the $Z'$ production
and its  decay into dileptons for the StSM and using
 $\sigma(pp\to G\to l^+l^-)$  for the production of the  graviton and its  decay into dileptons for the RS model.
       The dashed line is for the RS case with $k/{\bar{M}}_{Pl} = .01$. The allowed regions  are
        constructed by utilizing  the constrained parameter spaces of StSM \cite{Feldman:2006ce} and of the  RS model
\cite{Davoudiasl:1999jd,Davoudiasl:2000wi,Abazov:2005pi}.
        }
\label{cross_St_RS}
\end{figure}

In Fig. (\ref{widths}) we give a comparison of the  signature spaces
for the decay of the StSM $Z'$  and of  the RS  graviton  in the
warped geometry model  using the decay width-resonance mass plane.
The allowed regions (shaded) for the two models are exhibited, where
the unshaded regions correspond to constrained regions of the
parameter spaces of the two models. One finds that although there is
a region of the parameter space  of the RS model where the decay
widths can be  narrow,  the  region of potential overlap with  the
StSM is avoided if one includes the constrains  of  the oblique
parameters \cite{Peskin:1991sw,Altarelli:1991fk}. Fig.
(\ref{cross_St_RS})  gives a  more direct   method for
differentiating the two classes of models. Here one has plots of
$\sigma\cdot Br (Z'\to l^+l^-)$ and $\sigma\cdot Br (G\to l^+l^-)$
 as a function of the resonance mass. One finds  that the allowed  regions of the signature  space  of the two models
 consistent with the
 parameter space constraints  provides  a clear differentiation between these  two classes of models.
 Thus  Fig. (\ref{cross_St_RS}) provides an important tool for establishing the nature of the resonance once
 a  narrow resonance  is discovered.  Thus, for example, the
$\sigma\cdot Br (Z'\to l^+l^-)$ is an order of magnitude or more
smaller than $\sigma\cdot Br (G\to l^+l^-)$ over most of the
dilepton invariant mass that will be probed by the Drell-Yan process
at the  LHC.

\subsection{Angular Distributions in the Dilepton Channel in $pp\to (Z', G)\to  l^+l^-$}
\begin{figure}[h]
\centering
\includegraphics[width=12cm,height=10cm]{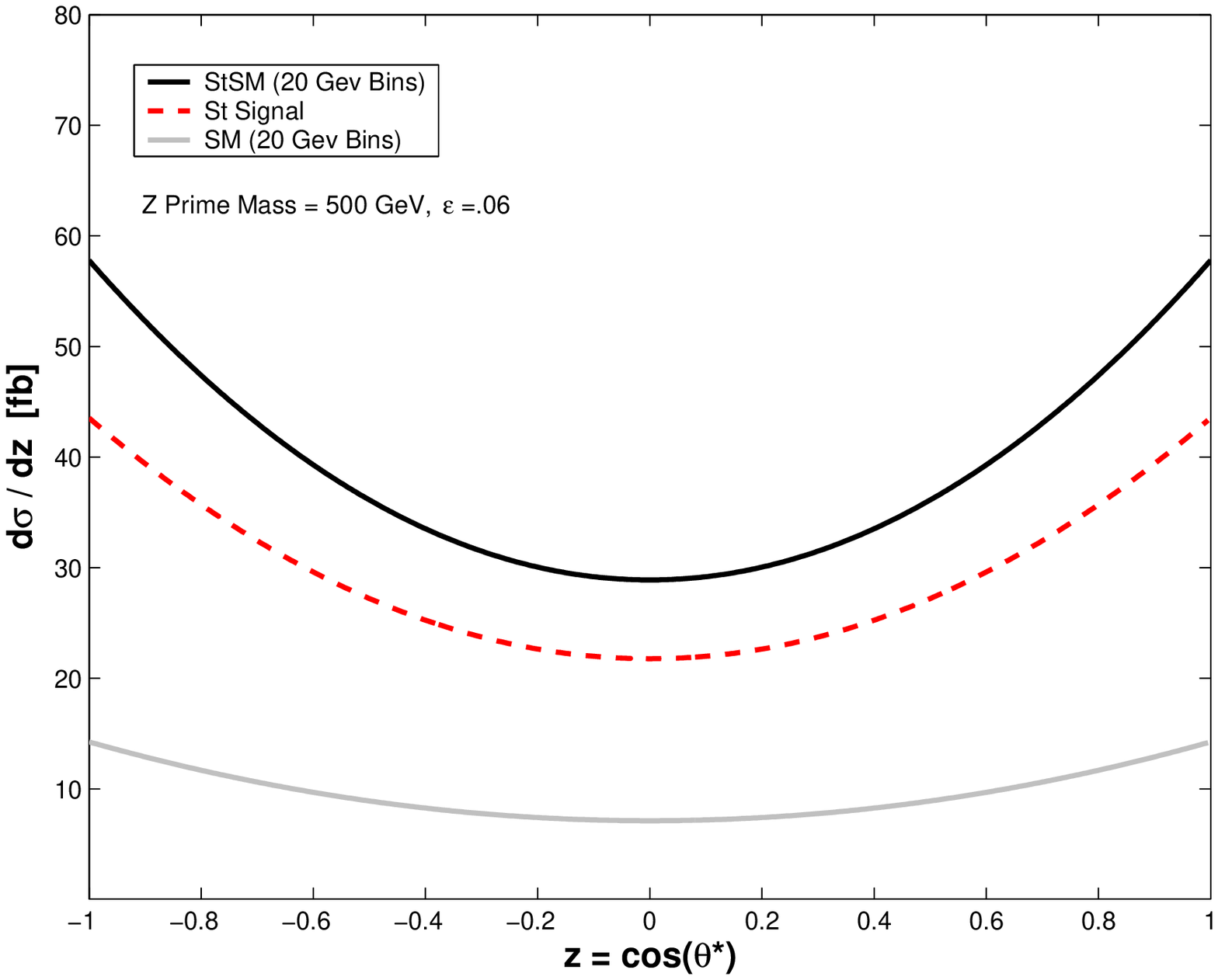}
\includegraphics[width=12cm,height=10cm]{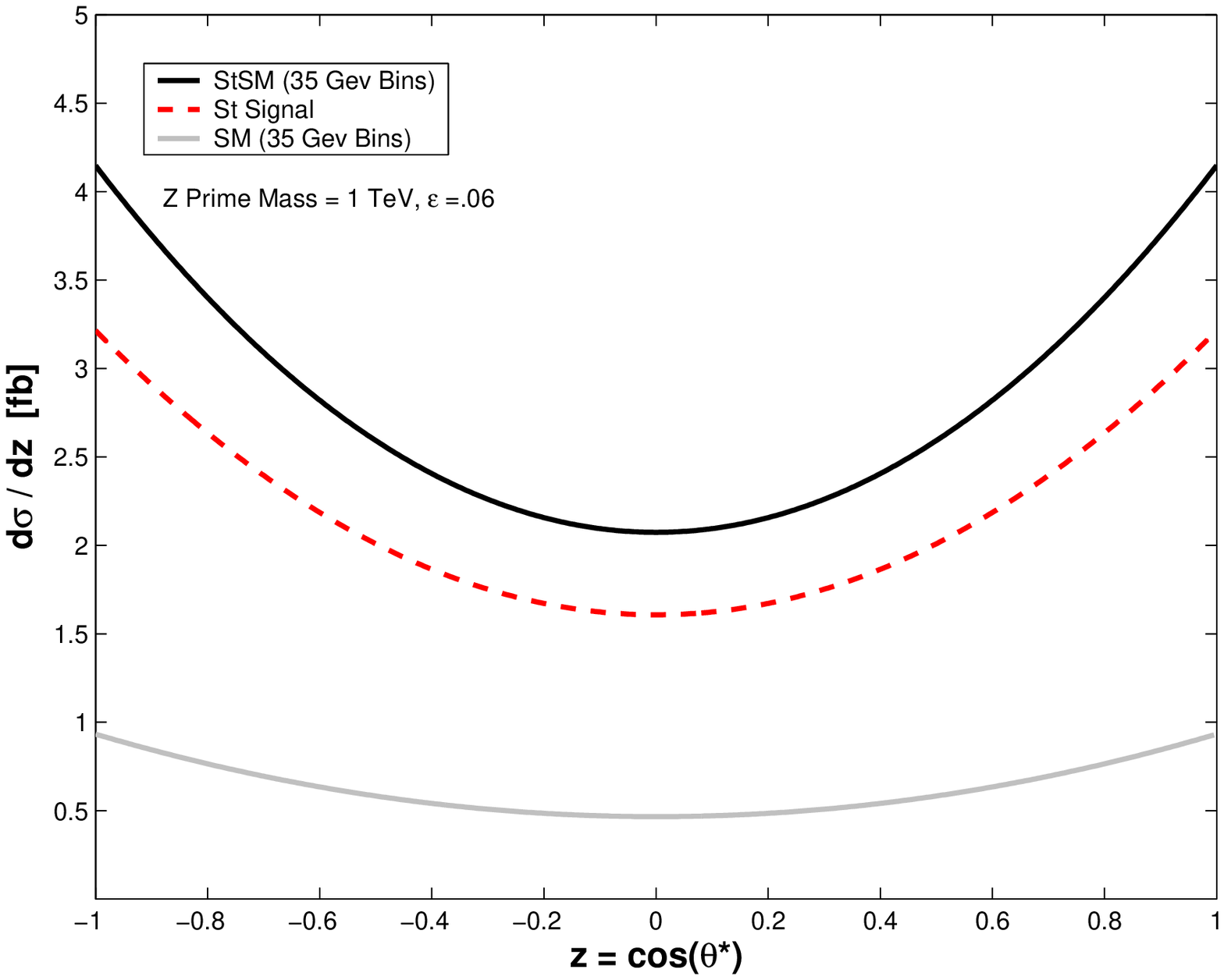}
\caption{ Angular distribution  $d\sigma/dz$ vs  $z=\cos(\theta^*)$
in the dilepton center of mass frame in the decay $Z'\to l^+ l^-$ in
StSM for $Z'$ mass of 500 GeV (upper graph) and 1 TeV (lower graph).
The SM background is also shown, and the StSM contribution sits high
above it.} \vspace{0.5in} \label{stsmang}
\end{figure}

Angular distributions in  the C-M frame of the final dilepton state
give clear signatures of the spin of the produced particle
 in the Drell-Yan process (for recent works see, for example,  Refs.\cite{Allanach:2002gn,Cousins:2005pq}).
  Thus angular distributions are  a powerful tool in distinguishing the
StSM $Z'$,   a spin 1 particle,  from the massive graviton of warped
geometry,  a spin 2 particle. The  CDF group has already carried out
angular distribution analyses \cite{Abulencia:2006iv}  using the
cumulative data at the Tevatron  and  more detailed analyses  are
likely to follow. Similar  analyses at the LHC would allow one to
investigate the spin of an observed resonance  with much more data.
In the following we give  a relative comparison of the angular
distributions arising from the StSM $Z'$ and from the massive
gravtion of warped geometry. To this end we first  examine the
feasibility of distinguishing the StSM $Z'$ signal from the Standard
Model background. This is done in   Fig. (\ref{stsmang})  for   $Z'$
masses  of  500 GeV and as well as 1 TeV with a bin size of 20 GeV
and 35 GeV respectively.  Fig. (\ref{stsmang}) shows  that the StSM
$Z'$ signal in this case is distinct from the $\gamma, Z$
background. Second, the StSM angular distribution sits high above
the SM background and thus an observation of  such a distribution
can lead to an identification of new physics in the dilepton
channel.

\begin{figure}[h]
\hspace*{-.2in} \centering
\includegraphics[width=12cm,height=10cm]{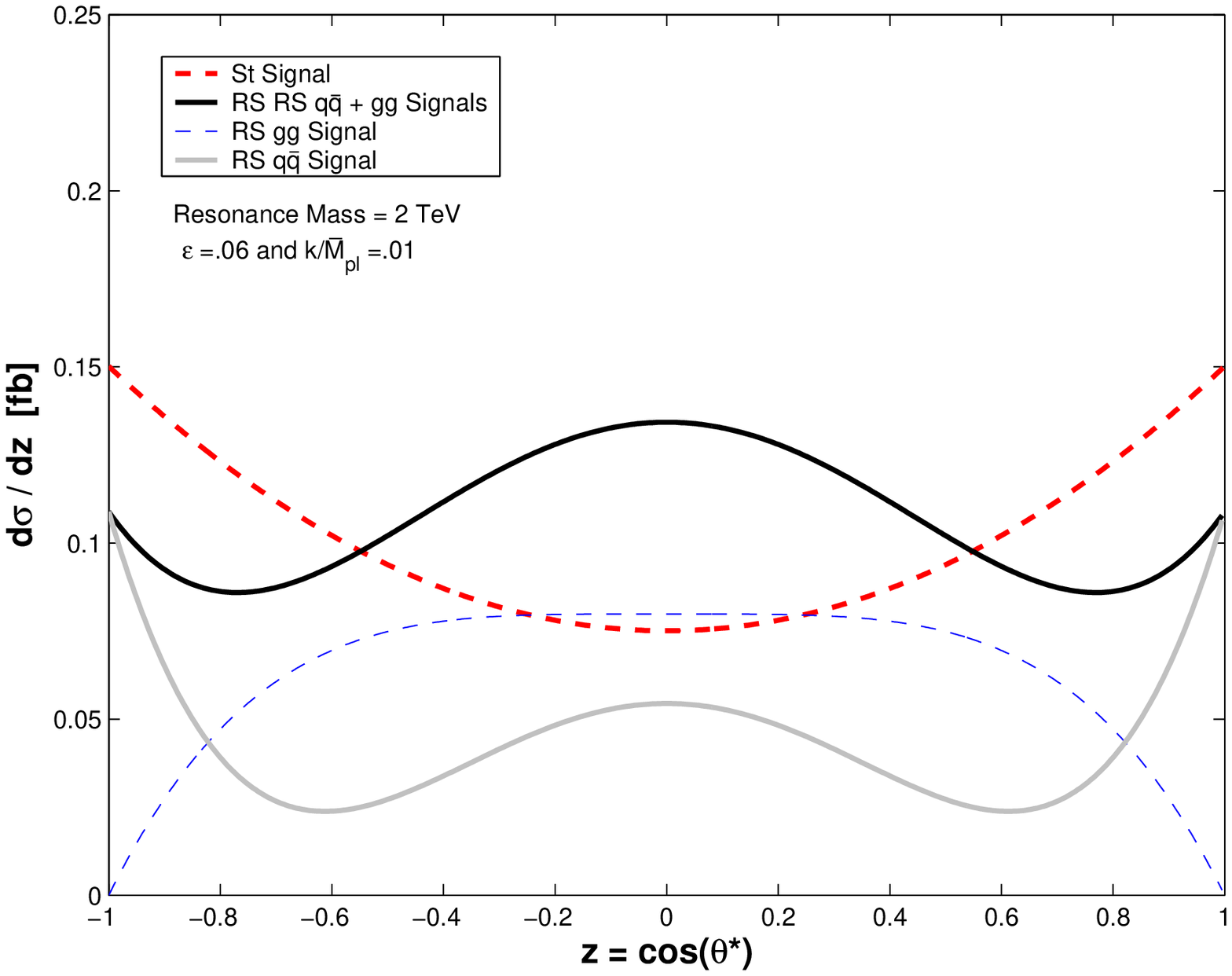}
\includegraphics[width=12cm,height=10cm]{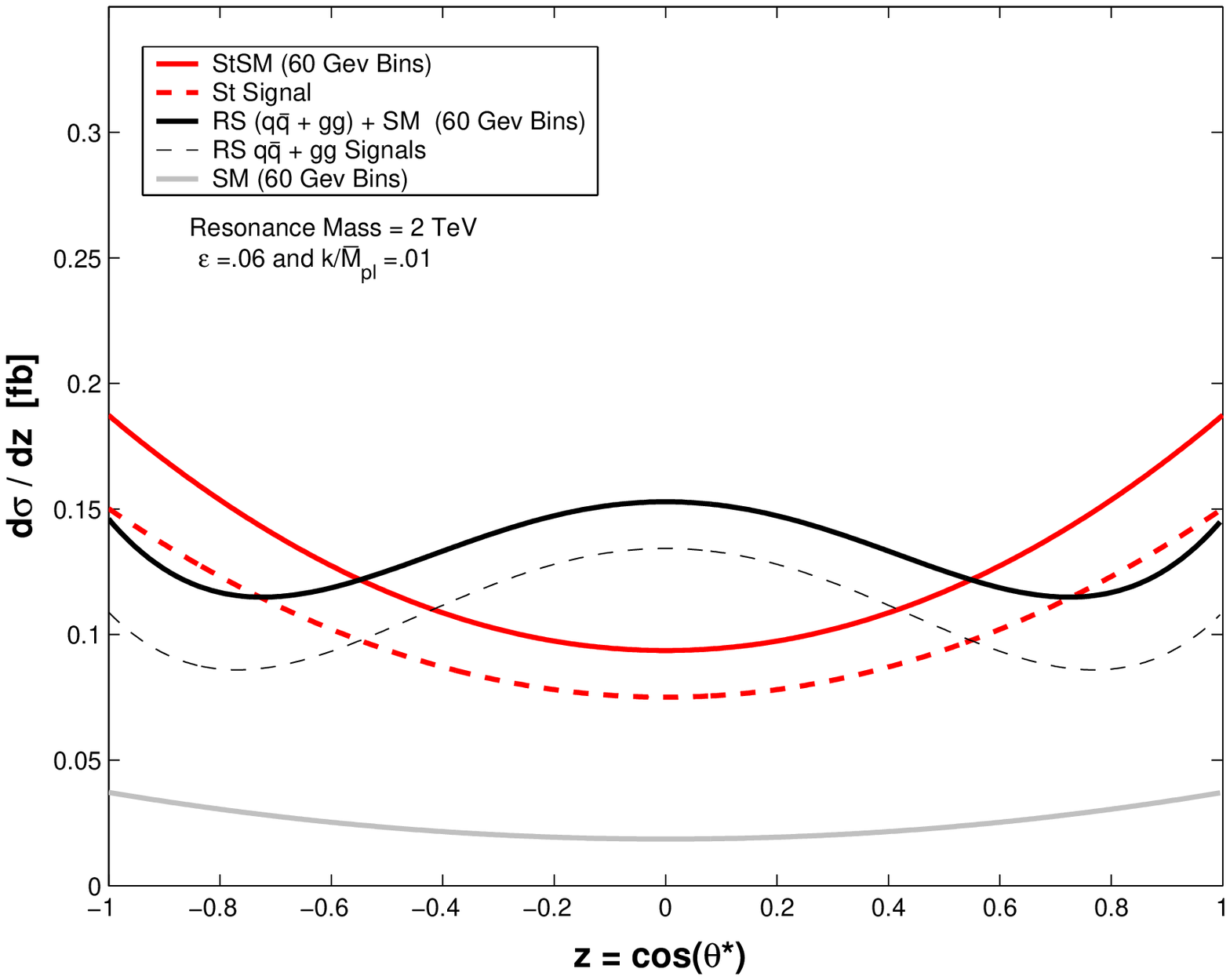}
\caption{An exhibition of the angular distribution $d\sigma(pp \to
Z'\to l^+l^-)/dz$ for the StSM model and  $d\sigma(pp \to G\to
l^+l^-)/dz$  for the RS model in the dilepton  center of mass
system, as defined  in Eqs. (\ref{Stz},\ref{RSz}). For the StSM,
$\epsilon$ is taken at .06 and $G$ is the first resonant mode of the
RS model, with $(k/\bar{M}_{Pl}) = .01$  and the resonance  mass is
2 TeV in each case.   For the RS model  the parameter  choice
requires  relaxing the  oblique constraints  and  the constraint on
$\Lambda_{\pi}$.} \vspace{0.5in} \label{RS-St-ang-Signals+SM}
\end{figure}

Next  we give a relative comparison of the  angular distribution in
the dilepton  channel arising from the StSM $Z'$ and the massive
graviton of warped geometry.  This is done in Fig.
(\ref{RS-St-ang-Signals+SM}) for a resonance mass of 2 TeV, the mass
region where an overlap between the two models can occur if the
constraints on the RS model are relaxed. The top graph in Fig.
(\ref{RS-St-ang-Signals+SM})  gives the angular  distributions
arising for the $Z'$  exchange but without the  Standard Model
background, i.e., what is plotted  is  the pure signal. Also plotted
is the pure signal from the graviton exchange which consists  of
contributions  from the quarks and  the gluons  which are separately
exhibited. In the lower graph of Fig. (\ref{RS-St-ang-Signals+SM})
the angular  distributions arising for the StSM $Z'$ and for the
massive graviton exchanges including  the Standard Model background
are  exhibited. The graph shows that the signal plus the background
lies significantly higher  than the SM background, and further the
sum of the $Z'$ signal and the SM background is easily
distinguishable from the sum of the massive graviton signal and the
SM background. The angular distributions for the graviton exchange
are  sensitively dependent on the graviton mass,  mainly due to the
sensitivity of the PDF \cite{Pumplin:2002vw} for the gluon  on the
mass scale. Thus the angular  distributions for the graviton will
change with the mass scale and change significantly.
 However, the angular  distributions for the $Z'$  and for  the graviton will continue to be
identifiably  distinct and allow one to distinguish between these
two classes  of narrow resonance models.

%
\section{Conclusions}

In this paper  we have carried out an investigation of narrow
resonances with specific focus on two classes of models  which have
recently emerged where narrow resonances arise  quite naturally. The
first of these are the Higgless  extensions of the Standard Model
gauge group, and of  the  Left-Right symmetric model gauge  group
 where the  extra gauge boson becomes
massive via the Stueckelberg mechanism.  A narrow $Z'$ naturally
arises  in these models.    The second class of models are those
based on warped geometry which give rise to a narrow graviton
resonance for $k/\bar M_{Pl}\sim .01$. The main focus of this paper
was to investigate the capability of the LHC to discover narrow
resonances specifically belonging to these classes of models and to
discriminate between them by examining their signature spaces.
  For the Stueckelberg model we  discussed the constraints on the
parameters space of the model using the LEP data and  the CDF and
D\O\ data.
  These constraints were  then
utilized to explore the narrow  Stueckelberg  $Z'$ at the LHC. The
analysis using the dilepton production  in the Drell-Yan process via
the $Z'$ boson shows that one will be able to explore a narrow $Z'$
resonance of Stueckelberg origin up to about 2 TeV with 100
${\rm{fb}}^{-1}$ of integrated luminosity and further up to 2.5 TeV
with 300 ${\rm{fb}}^{-1}$ of integrated luminosity.  With 1000
${\rm{fb}}^{-1}$ of integrated luminosity one could even explore a
Stueckelberg $Z'$   beyond 3 TeV. The results of this analysis are
summarized in Fig. (\ref{searchreachepslionandwidth2}) and  Fig.
(\ref{bars}).

We carried out a similar analysis for the dilepton production in the
warped geometry RS model which also has the potential of supporting
a narrow resonance. It is then interesting to ask how a Stueckelberg
type narrow resonance could be  distinguished from a narrow massive
graviton of warped geometry. Indeed  there is a range of the
parameter space where an overlap exists between the two  models with
the width of the massive graviton of the warped geometry being
similar to the width of the $Z'$  arising from the Stueckelberg
model. We have shown that one of the clear distinguishing features
between them is $\sigma \cdot Br(l^+l^-)$ for dilepton production in
the Drell-Yan process which proceeds through the interaction $pp\to
Z'\to l^+l^-$ for the Stueckelberg model and via $pp\to G \to
l^+l^-$ for the case of the RS model.  The analysis of Fig. (\ref
{cross_St_RS})  shows that for any resonance mass the signature
spaces of the StSM and of the RS model are distinct and one can
discriminate between them using   the  $\sigma \cdot Br (l^+l^-)$
criterion. In addition,  the angular  distributions in the dilepton
center of mass system provide a clear discrimination between the two
models. Here one finds that the angular distributions from the  StSM
$Z'$ and from the massive graviton lie well above the Standard Model
background and further are  distinctly dissimilar as exhibited in
the analysis of Fig. (\ref{RS-St-ang-Signals+SM}).

\begin{figure}[htb]
  \begin{center}
  \epsfig{figure=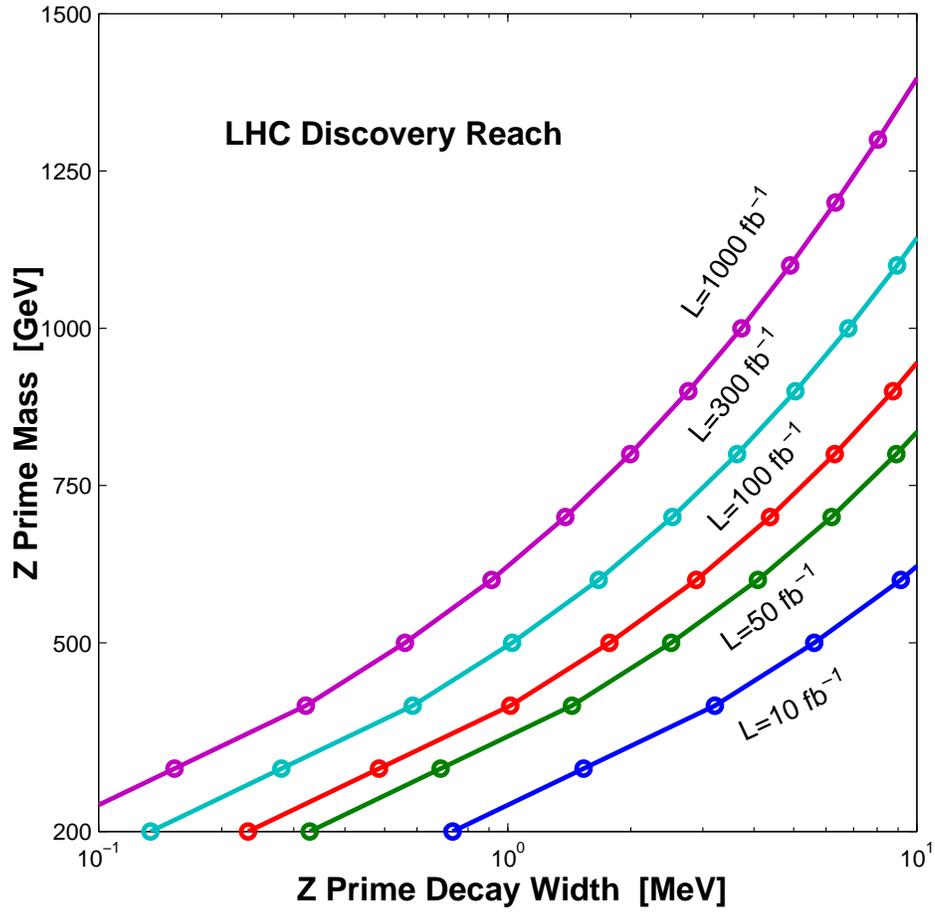,width=.85\textwidth}
\caption{ A plot of the discovery reach of the LHC for small StSM
$Z'$ widths. The allowed  regions are  to the right and below each
curve for a given luminosity. This figure is  a blown up version of
the very low  $Z'$  width region of  Fig
(\ref{searchreachepslionandwidth2}).} \label{submevwidth}
  \end{center}
\end{figure}

Some general features of the searches for narrow resonances  were
also discussed. The bin size used in data collection has  a direct
bearing on the signal to background ratio as shown in Fig.
(\ref{signoise}). The analysis presented  in this paper  reveals the
remarkable phenomenon that the models considered  here  can be
tested even when the resonance widths are small and the resonance
masses are large. Specifically one finds that the StSM model can
produce observable cross section signals with a $Z'$ width lying in
the  MeV or even in the sub-MeV range while the $Z'$ mass may be in
hundreds of GeV to TeV  range. This phenomenon is exhibited  in
Fig. (\ref{submevwidth}).  While the result of
  Fig. (\ref{submevwidth})  is presented for the specific case of
StSM $Z'$ model,
 similar considerations may apply to a  wider  class of models
which support a
  narrow resonance.
   The evidence for a narrow resonance  will be  an
important hint
  for an altogether new  type of physics  beyond the  Standard Model
  and possibly a hint of a string origin.\\

\section*{Acknowledgements}
We  thank George Alverson, Emanuela Barberis,  and  especially
Darien Wood for many informative discussions relating to experiment.
We thank Greg Landsberg for a communication  regarding the  analyses
of the D\O\  Collaboration, and thank  Ben Allanach, and  the
Cambridge SUSY Working Group, for a communication regarding the
detector and acceptance/efficiency cuts used in their work on the
LHC analysis. This research  was supported in part by NSF grant
PHY-0546568.
\clearpage

\clearpage

%
\begin{table}[h]
\vspace{3 cm} \centering
 \title{\bf StSM Electroweak Fit}\vspace{.3 cm}
\begin{tabular}{||c||c||c||c||c||c||}
\hline\hline
Quantity                       & Value  (Exp.)                     &StSM        &$\Delta$Pull        \\
\hline
$\Gamma_Z$ [GeV]           & 2.4952  $\pm$ 0.0023      &(2.4952-2.4942)       &(0.2, 0.6)        \\
$\sigma_{had}$ [nb]        & 41.541  $\pm$ 0.037       &(41.547-41.568)       &(-0.3, -0.9)      \\
$R_e$                     & 20.804  $\pm$ 0.050       &(20.753-20.761)       &(-0.1, -0.2)        \\
$R_\mu$                    & 20.785  $\pm$ 0.033       &(20.800-20.761)       &(-0.1, -0.4)        \\
$R_\tau$                   & 20.764  $\pm$ 0.045       &(20.791-20.807)       &(-0.1, -0.3)        \\
$R_b$                      & 0.21643 $\pm$ 0.00072     &(0.21575-0.21573)     &(0.0, 0.0)        \\
$R_c$                      & 0.1686 $\pm$ 0.0047       &(0.1711-0.1712)       &(0.0, 0.0)        \\
$A^{(0,e)}_{FB}$           & 0.0145 $\pm$ 0.0025       &(0.0168-0.0175)       &(-0.2, -0.5)      \\
$A^{(0,\mu)}_{FB}$         & 0.0169  $\pm$ 0.0013      &(0.0168-0.0175)       &(-0.3, -0.9)      \\
$A^{(0,\tau)}_{FB}$        & 0.0188  $\pm$ 0.0017      &(0.0168-0.0175)       &(-0.2, -0.7)      \\
$A^{(0,b)}_{FB}$           & 0.0991  $\pm$ 0.0016      &(0.1045-0.1070)       &(-0.8, -2.3)     \\
$A^{(0,c)}_{FB}$           & 0.0708  $\pm$ 0.0035      &(0.0748-0.0766)       &(-0.3, -0.8)      \\
$A^{(0,s)}_{FB}$           & 0.098  $\pm$ 0.011      &0.105-0.107)         &(-0.1, -0.3)      \\
$A_e$                      & 0.1515 $\pm$ 0.0019       &(0.1491-0.1524)       &(-1.0, -2.8)      \\
$A_\mu$                    & 0.142  $\pm$ 0.015        &(0.149-0.152)         &(-0.1, -0.4)      \\
$A_\tau$                   & 0.143  $\pm$ 0.004        &(0.149-0.152)         &(-0.5, -1.3)      \\
$A_b$                      & 0.923  $\pm$ 0.020        &(0.935-0.935)         &(0.0, 0.0)        \\
$A_c$                      & 0.671  $\pm$ 0.027        &(0.669-0.670)         &(0.0, 0.1)        \\
$A_s$                      & 0.895 $\pm$ 0.091         &(0.936-0.936)         &(0.0, 0.0)  \\
\hline\hline
\end{tabular}
\caption{\label{tab:table1} Results of the StSM fit to a standard
set of electroweak observables  at the $Z$ pole for $\epsilon$ in
the range  $(.035-.059)$ for $M_1=350$ GeV. The Pulls are calculated
as shifts from the SM fit via $\Delta {\rm Pull}=({\rm SM -
StSM})/\delta {\rm Exp}$ and
  Pull(StSM)=Pull(SM)+ $\Delta$Pull.
 The  data in column 2 is taken from  Ref. \cite{pdg}.}
  \end{table}
  \clearpage

\begin{table}[h]
\centering
 \title{\bf Comparing the StSM and StLR}\vspace{.3 cm}
\begin{tabular}{||c||c||c||c||c||}
\hline\hline
Quantity                       & StSM                 &StLR   \\
\hline\hline
$\epsilon =M_2/M_1$     &.060          & .071 \\
$M_{Z'} $  [GeV]            &500         &500  \\
($v'_{\nu}$ , $a'_{\nu}$)      &(0.014638, 0.014638)        &(0.014615, 0.014621)           \\
($v'_e$ , $a'_e$)                 & (0.042401, -0.014638)         &(0.042352, -0.014621)              \\
($v'_u$ , $a'_u$)                   &(-0.023388, 0.014638)       &(-0.023363, 0.014621)            \\
($v'_d$ , $a'_d$ )                & (0.004375, -0.014638)        &(0.004374, -0.014621)             \\
$\Gamma_Z$ [GeV]        & 0.0297       & 0.0299             \\
$Br(\nu_e\bar\nu_e)$        & 2.36\%       &2.60\%         \\
$Br(e^+e^-)$                       &12.33\%          &12.33\%              \\
$Br(u\bar u)$                       & 14.52\%         &14.42\%                 \\
$Br(d\bar d)$                     & 4.45\%         &4.42\%                 \\
$Br(t\bar t)$                      &10.93\%          &10.85\%                \\
$Br(W^+W^-)$                  &2.60\%        &2.56\%                 \\
    \hline\hline
\end{tabular}
\caption{\label{stsmandstlr} Comparison of  the $Z'$ branching
ratios in StSM and StLR model at $M_{Z'}=500$ GeV for the maximum
allowed value of  $\epsilon$ consistent with the analysis of Sec.
(\ref{ConstZmass}). The couplings and branching ratios for the $Z'$
in the two  models turn out be remarkably close. }
  \end{table}

\begin{table}[h]
\centering
 \title{\bf Production Cross Sections in the  StSM and RS Models}\vspace{.3 cm}
\begin{tabular}{||c||c||c||c||c||}
\hline\hline $(M_{Z'} , M_G)$ & $\Gamma_{Z'} $  (GeV) & $\Gamma_{G}$
(GeV) &  $\sigma_{Z'}\cdot Br$ (fb)&$\sigma_{G}\cdot Br $ (fb)
\\\hline
 1000  & 0.058& 0.141       & 4.29     & 9.98 \\
1250  & 0.073& 0.176       & 1.72  & 3.11       \\
1500  &  0.087& 0.212       & 0.779 & 1.15     \\
1750   &  0.102& 0.247    & 0.384 & 0.475      \\
2000 & 0.117 & 0.283     & 0.200 & 0.215      \\
2250 & 0.131 & 0.318        & 0.109 & 0.104     \\
2500   & 0.146 & 0.354     & 0.061 & 0.053      \\
2750 & 0.160& 0.389       & 0.035  & 0.028   \\
3000  & 0.175 & 0.425       & 0.021& 0.015   \\
\hline \hline
\end{tabular}
\caption{\label{tab:table2}\small A comparison of the narrow
resonance widths  and $\sigma.Br(l^+l^-)$ in StSM  for $\epsilon =
.06$ and in the RS warped geometry with $k/ {\bar M}_{Pl} = .01$ as
a function of the resonance mass in GeV. }
\end{table}

\begin{table}[h]
\begin{center}
 \title{\bf Events in the  StSM and RS Models}\vspace{.5cm}
\begin{tabular}{||c||c||c||c|c||}
\hline\hline   $(M_{Z'} , M_G)$ & Bin (GeV) & $N_{SM}$ &
$N_S =(N_{St},N_{RS})$ & $N^{min}_S$ \\
\hline\hline
1000 & 30.65 &54.45 & (214.33,716.96)    & 36.90 \\
1250 & 36.79 & 20.95  & (85.90,216.96)   & 22.89     \\
1500 & 42.89 & 9.22 & (38.94,77.73)     & 15.18     \\
1750 &  48.96& 4.44 & (19.18,31.30)     & 10.53    \\
2000 & 55.02 & 2.27 & (10.01,13.72)    & 10   \\
2250 &  61.07& 1.22 & (5.46,6.41)     & 10   \\
2500 &  67.11& 0.68 & (3.07,3.15)       &10   \\
2750 &73.14  &0.39  & (1.77,1.60)     &10   \\
3000 & 79.17 & 0.22  & (1.04,0.84)  & 10    \\
\hline\hline
\end{tabular}
 \end{center}
\caption{\label{tab:table3}\small A comparison  of the signal events
with  integrated luminosity of $\mathcal L =$ 100 $\rm fb^{-1}$  in
the StSM for the case $\epsilon = .06$  with the  signal in the RS
warped geometry for $ k/ {\bar M}_{Pl} = .01$ including ATLAS
detector effects as a function of the resonance mass in GeV.
 Acceptance($A$) and efficiency($\varepsilon$)  for
the RS case is as in  Ref. \cite{Allanach:2000nr}, while for the
StSM we use the spin 1 detector losses given in Ref. \cite{SLT}
$\approx$ 50 \% as discussed in the text. For $X=(Z',G)$ of Table
\ref{tab:table2}, $N_S = (\sigma\cdot Br)\varepsilon  A \mathcal L$,
$N_B =N_{SM}$ (background integrated over the bin), $N_S^{min}=
5\sqrt{N_B }$ or 10, whichever is larger. The minimum signal cross
section is $ (\sigma\cdot Br)^\mathrm{min } = (\varepsilon A
\mathcal L)^{-1} N_S^{min}$ for each model. }\end{table}


\clearpage

%

%


%
%



\end{document}